\documentclass[pdflatex,sn-mathphys-num]{sn-jnl}

\usepackage{graphicx}%
\usepackage{multirow}%
\usepackage{amsmath,amssymb,amsfonts}%
\usepackage{amsthm}%
\usepackage{mathrsfs}%
\usepackage[title]{appendix}%
\usepackage{xcolor}%
\usepackage{textcomp}%
\usepackage{manyfoot}%
\usepackage{booktabs}%
\usepackage{algorithm}%
\usepackage{algorithmicx}%
\usepackage{algpseudocode}%
\usepackage{listings}%
\usepackage{graphicx,color}
\usepackage{epsfig}
\usepackage{dcolumn}
\usepackage{amsmath,amssymb,bm}
\usepackage{amsmath}
\usepackage{hyperref,url}
\usepackage[noabbrev]{cleveref}
\usepackage{CJK}
\usepackage{lineno}
\usepackage[english]{babel}
\usepackage[T1]{fontenc}
\usepackage{pifont}
\usepackage{amsfonts}
\usepackage{wasysym}
\usepackage{amsbsy}
\usepackage{psfrag}
\usepackage{color}
\usepackage{multirow}
\usepackage{cases}
\usepackage{stackengine}
\usepackage{float}
\usepackage{blkarray}
\usepackage{cancel}
\usepackage{mathrsfs}
\usepackage{empheq}
\usepackage{amssymb}
\usepackage{upgreek}
\usepackage{amsmath}
\usepackage{etoolbox}
\usepackage{comment}
\usepackage[version=4,arrows=pgf-filled,
    textfontname=sffamily,mathfontname=mathsf]{mhchem}
\usepackage{physics}
\usepackage{bm}

\theoremstyle{thmstyleone}%
\theoremstyle{thmstyletwo}%

\theoremstyle{thmstylethree}%

\begin{document}

\title[Article Title]{%
  {\bfseries\centering Biogenic bubbles enable microbial escape \\ from physical confinement}%
}

\author[1]{\fnm{Babak} \sur{Vajdi Hokmabad}}

\author[2]{\fnm{Thomas} \sur{Appleford}}

\author[1]{\fnm{Hao} \sur{Nghi Luu}}

\author[3,4]{\fnm{Meera} \sur{Ramaswamy}}

\author[2,5]{\fnm{Maziyar} \sur{Jalaal}}

\author*[6,1]{\fnm{Sujit S.} \sur{Datta}} \email{ssdatta@caltech.edu}

\affil[1]{\centering\orgdiv{Department of Chemical and Biological Engineering},\\ \orgname{Princeton University}, \city{Princeton},  \state{NJ}, \country{USA}}

\affil[2]{\centering\orgdiv{Van der Waals-Zeeman Institute, Institute of Physics}, \\\orgname{University of
Amsterdam}, \city{Amsterdam}, \country{The Netherlands}}

\affil[3]{\centering\orgdiv{Department of Mechanical Engineering}, \\\orgname{University of Minnesota}, \city{Minneapolis},  \state{MN}, \country{USA}} 

\affil[4]{\centering\orgdiv{Princeton Center for Complex Materials}, \\\orgname{Princeton University}, \city{Princeton},  \state{NJ}, \country{USA}} 

\affil[5]{\centering\orgdiv{Department of Applied Mathematics and Theoretical Physics}, \\\orgname{University of Cambridge}, \city{Cambridge}, \country{UK}} 

\affil[6]{\centering\orgdiv{Division of Chemistry and Chemical Engineering}, \\\orgname{California Institute of Technology}, \city{Pasadena},  \state{CA}, \country{USA}}

\abstract{Immotile microbes inhabit nearly every environment on Earth, from soils and sediments to food matrices---yet how they disperse through these physically confining environments is poorly understood. Here, we show that immotile microbial colonies confined in a model transparent yield-stress matrix can achieve long-range dispersal by harnessing their own metabolism. Using yeast as a model organism, we find that fermentation drives dissolved CO$_2$ to supersaturation, nucleating biogenic bubbles that grow, yield the matrix, and rise, hydrodynamically entraining cells vertically in their wake. Sequential bubble nucleation sculpts persistent columnar colonies extending far beyond what growth alone permits. Multiple colonies interact via their fermentation byproducts, merging and mixing genetically as they collectively sculpt self-sustaining conduit networks. Our findings reveal a third mode of microbial dispersal, distinct from the canonical mechanisms of motility and growth, with implications for ecology, environmental science, and biotechnology. More broadly, they exemplify a previously unrecognized class of active behavior---Metabolically Driven Active Matter---in which metabolic byproducts reshape the physical landscape of confinement to drive population-scale motion.}


\maketitle

Microbes typically inhabit confining environments, from soils and sediments~\cite{hayat2010soil,philippot2024interplay,bhattacharjee2019bacterial, young2004interactions} to the food we eat~\cite{bokulich2016new}.
The ability to disperse through these environments is fundamental to microbial life: it enables cells to find nutrients, access new ecological niches, and sustain the biogeochemical cycling that underpins global carbon and nutrient fluxes~\cite{tao2023microbial, peng2022wetland, cavicchioli2019scientists,thauer2008methanogenic}. Dispersal is canonically understood through the lens of motility, the ability of cells to self-propel~\cite{bhattacharjee2019bacterial, cremer2019chemotaxis}. Yet, many microbes are immotile~\cite{vanwonterghem2016methylotrophic}. Growth-driven colony expansion is widely assumed to be the only remaining dispersal route for these microbes, but it is inherently slow; as a colony grows, nutrient consumption outpaces influx, driving a transition from exponential to surface-limited growth~\cite{martinez2022morphological,hallatschek2023proliferating,lavrentovich2013nutrient, kannan2025spatiotemporal}. Immotile microbes are therefore thought to remain locally confined. How, then, have they colonized nearly every ecological niche in the biosphere~\cite{fyfe1996biosphere, colman2017deep}?

\begin{figure}[h]
\centering
\includegraphics[width=0.95\textwidth]{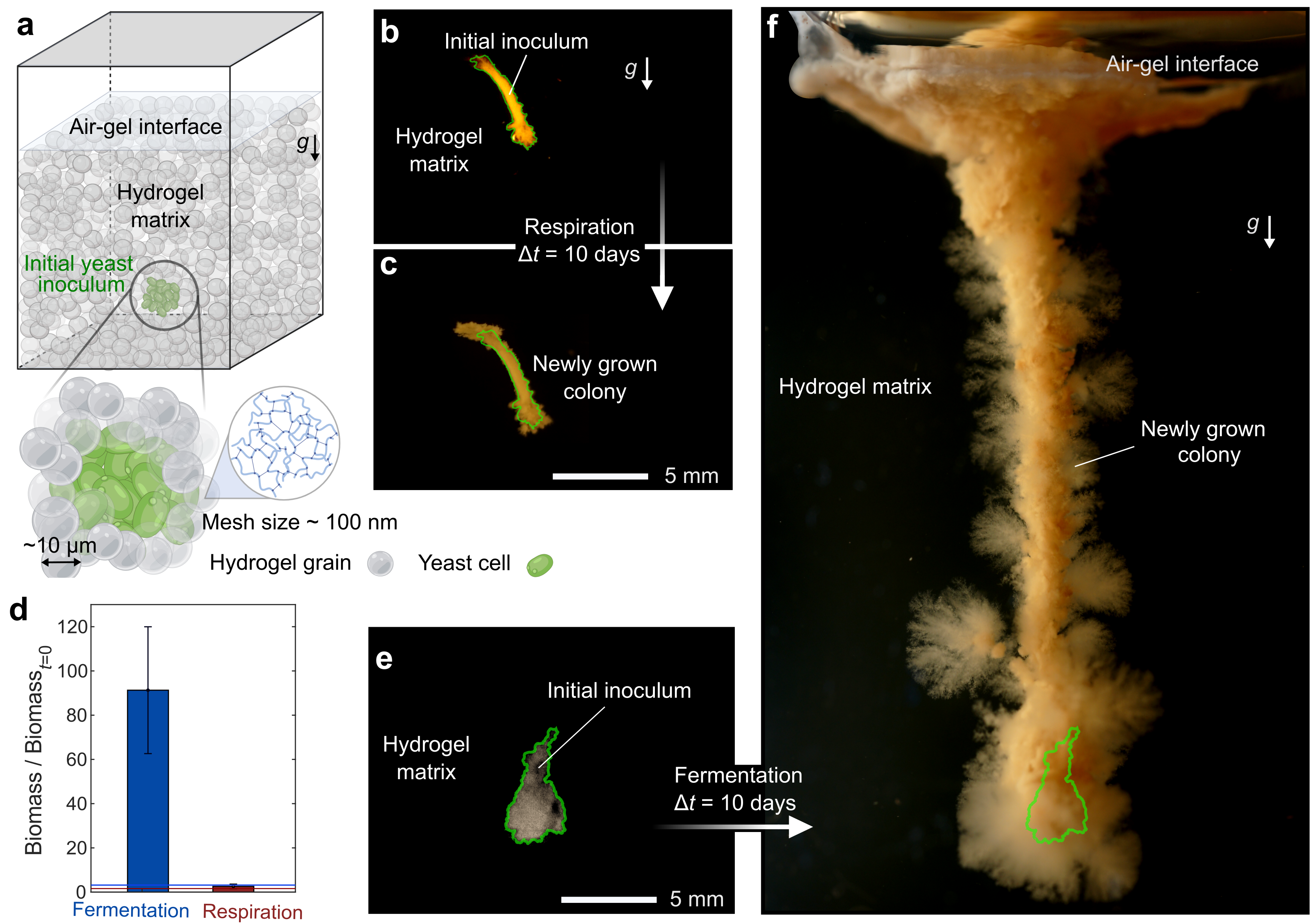}
\caption{\textbf{Fermentation drives long-range vertical dispersal of confined yeast colonies.} ~\textbf{a,} Experimental setup. The matrix is a jammed packing of transparent hydrogel grains (diameter $\sim10 ~\upmu \rm m$) swollen in yeast growth medium, sealed within a transparent chamber; each grain is permeable to fluid and dissolved gases/nutrients. We inject a dense colony of yeast into the matrix, avoiding residual mechanical strain or cell trails to ensure unbiased initial conditions, and image its dynamics using a digital camera from the side. \textbf{b,} Initial inoculum and \textbf{c}, final colony after 10 days in YPG growth medium with $2\ \rm w/v\% $ glycerol as the primary carbon source (non-fermentable). The colony remains locally confined, expanding slightly with a rough branching morphology characteristic of surface-limited growth. \textbf{d,} Comparison of biomass production after 10 days under fermentation (dextrose) and respiration (glycerol) conditions. Solid lines show the theoretical prediction for a spherical colony under surface-limited growth. Fermentation produces a $\sim90\times$ increase in biomass, far exceeding both the respiration case and predictions for surface-limited growth (solid lines), demonstrating that fermentation enables a qualitatively distinct dispersal mode. \textbf{e,} Initial inoculum and \textbf{f,} final colony after 10 days in YPD growth medium with $2\ \rm w/v\% $ dextrose as the primary carbon source (fermentable). The colony grows into a vertically elongated column reaching the air–matrix interface $\sim4$~cm above the initial inoculum. In both \textbf{b-c} and \textbf{e-f}, the hydrogel concentration is $c=1  \text{ w/v}\%$, corresponding to a matrix yield stress $\sigma_{y}=10.7 ~\rm Pa$.}\label{fig:fig1}
\end{figure}

Here, we uncover a previously unknown strategy by which immotile microbes disperse over long distances through confining environments. We use baker's yeast (\emph{Saccharomyces cerevisiae}) embedded in optically transparent granular hydrogel matrices [Fig.~\ref{fig:fig1}\textbf{a}] as a model system---enabling direct visualization of microbial dynamics invisible in opaque natural environments. Our experiments reveal that microbial fermentation of sugars, a ubiquitous metabolic pathway, drives dissolved CO$_2$ to supersaturation, nucleating bubbles. As each bubble grows, it deforms and eventually yields the surrounding matrix, causing the bubble to rise and hydrodynamically entrain cells in its wake. Sequential bubble migration sculpts persistent columnar colonies extending vertically more than $40\times$ beyond what surface-limited growth alone permits, with a width governed not by biology, but by the competition between capillary and gravitational forces on the bubbles. These findings challenge a central assumption of microbial biophysics: that motility is required for long-range dispersal through confining environments. Instead, they demonstrate that microbes can harness their own metabolism---coupling chemical activity to the rheology and fluid mechanics of their surroundings---to disperse over distances far exceeding what growth alone permits.

We initialize each experiment by introducing a small inoculum of densely packed yeast (effective radius $R_{\rm colony} \sim 1$ mm, concentration $\textit{c}_{\rm cell} \sim 10^{12}$ cells mL$^{-1}$) into a granular hydrogel matrix [Fig.~\ref{fig:fig1}\textbf{a}, detailed in \emph{Supplementary Information}]. Each matrix is composed of biocompatible hydrogel grains swollen in a defined nutrient-rich liquid growth medium. The individual grains are freely permeable to nutrient and fluid, but are packed so densely that cells are physically confined in the pores between adjacent grains~\cite{martinez2022morphological,bhattacharjee2018polyelectrolyte,hancock2024interplay,hancock2025nutrient,bay20243d}---mimicking the structure of many natural microbial habitats. Moreover, similar to natural habitats, the matrices are yield-stress materials~\cite{bhattacharjee2018polyelectrolyte,lee2020migration} [Fig.~S1] with rheological properties tuned so they do not strongly constrain colony growth but rather keep the cells suspended in three dimensions (3D, yield stress $\sigma_y\sim 3-30$~Pa). Because the hydrogel grains are so highly swollen, the matrices are transparent, enabling us to directly visualize colony morphodynamics in situ. 

First, we explore the case of colony growth in a matrix with $2\ \rm w/v\%$ glycerol as the primary carbon source, promoting aerobic respiration. Due to consumption by the cells, we expect that nutrient penetration into the colony is limited to a depth $\approx\sqrt{D_{n}c_{n}/( k_{n}c_{\text{cell}})} \approx 20 ~\upmu \text{m}$ from the surface, spanning just a few cells, where $D_{n}$ is the nutrient diffusivity, $c_{n}$ is the the characteristic Michaelis-Menten nutrient concentration, and $k_n$ is the maximal consumption rate per cell~\cite{martinez2022morphological}. Thus, we expect that growth is limited to a thin surface layer of cells. Our experiments confirm this expectation: as shown in Fig.~\ref{fig:fig1}\textbf{b–c}, the colony remains confined, expanding slightly at its surface with a rough branching morphology characteristic of surface-limited growth~\cite{martinez2022morphological}. This expansion corresponds to a slight $\approx3\times$ increase in biomass [purple bar in Fig.~\ref{fig:fig1}\textbf{d}] after 10 days, again consistent with surface-limited growth [horizontal purple line in Fig.~\ref{fig:fig1}\textbf{d}, \emph{Supplementary Information}].

In nature, however, confined microbial colonies typically inhabit anaerobic conditions---under which fermentation is the dominant metabolic pathway for sugar consumption~\cite{brune2000life, jo2022gradients}. Therefore, we repeat the same experiment, but replacing the glycerol with $2\ \rm w/v\%$ dextrose as the primary carbon source, which the cells consume via fermentation instead. Surprisingly, in this case, the colony grows into a vertically elongated column that reaches the air–matrix interface $\sim4$~cm ($\sim4,000$ cell diameters) above the initial inoculum after 10 days [Fig.~\ref{fig:fig1}\textbf{e}]. This vertical expansion corresponds to a $\approx90\times$ increase in biomass [blue bar in Fig.~\ref{fig:fig1}\textbf{d}], far exceeding the increase expected from purely surface-limited growth alone [horizontal blue line in Fig.~\ref{fig:fig1}\textbf{d}, \emph{Supplementary Information}].

\begin{figure}[h]
\centering
\includegraphics[width=0.55\textwidth]{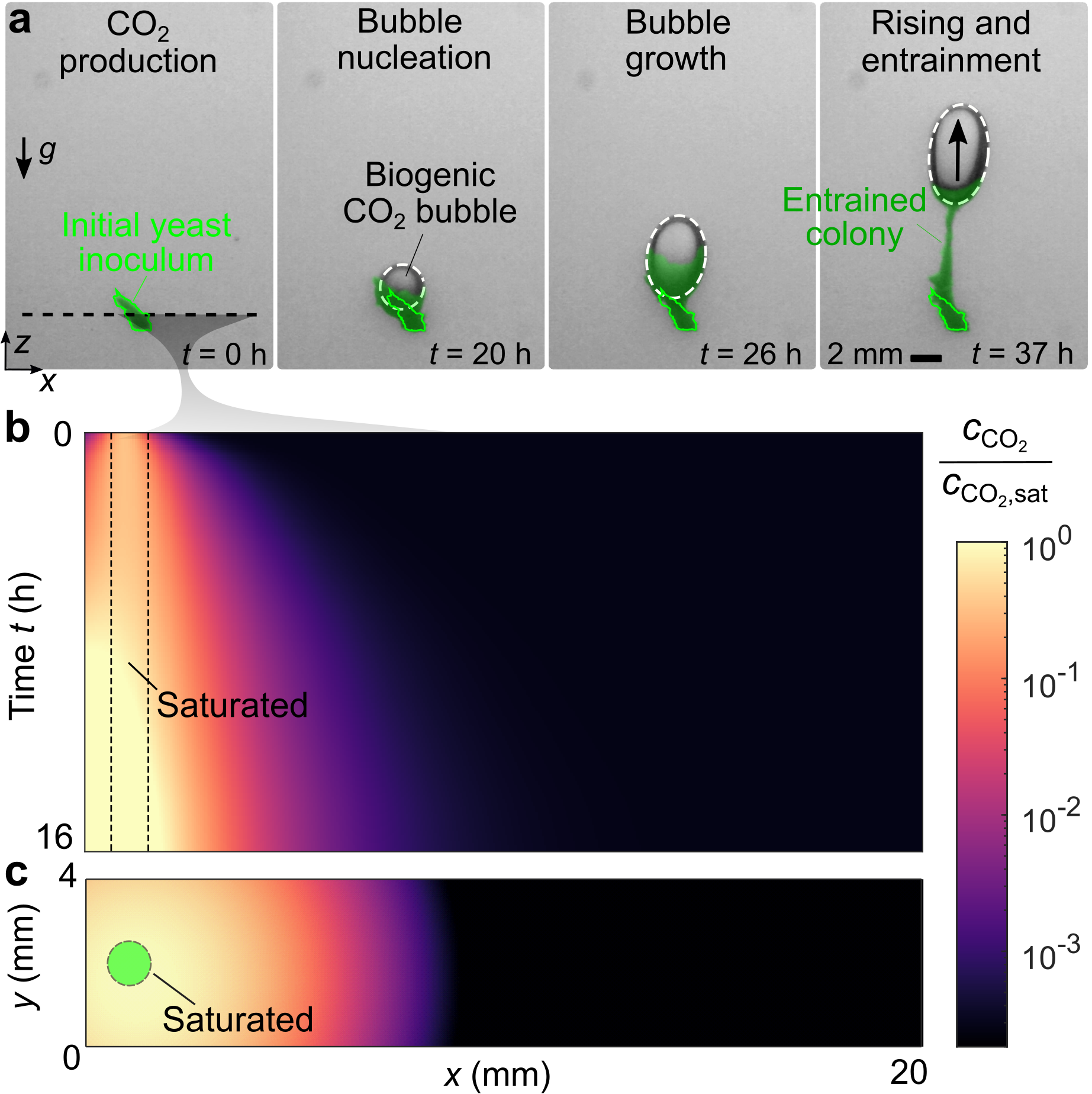}
\caption{\textbf{Biogenic CO$_{2}$ bubbles nucleate, grow, and vertically entrain the colony.} \textbf{a,} Shadowgraphy time series of a yeast colony undergoing fermentation in a hydrogel matrix ($c_{h}=0.9 \% \text{ w/v}$, $\sigma_y=7.7$ Pa, YPD medium). False color (green) delineates the colony at each time point; the brighter green outline marks the initial inoculum perimeter. A biogenic bubble nucleates at $t=20$~h, grows and elongates vertically ($t=26$~h), then detaches and rises, entraining a vertical column of cells in its wake ($t=37$~h). \textbf{b,} Kymograph of dissolved CO$_{2}$ concentration, $c_{\text{CO}_2}$, measured using sodium fluorescein (a pH-sensitive fluorophore) via confocal microscopy at the horizontal plane indicated by the dashed line in \textbf{a}. CO$_{2}$ produced by fermentation accumulates in the surrounding fluid, reaching saturation ($c_{\text{CO}_{2},\text{sat}}$) and triggering bubble nucleation. \textbf{c,} Spatial map of $c_{\text{CO}_2}$ at $t=16$~h, showing the buildup of dissolved CO$_2$ around the colony prior to nucleation. }\label{fig:fig2}
\end{figure}

What drives this transition from confined, surface-limited growth to long-range vertical dispersal? 
Direct visualization of the colony morphodynamics provides a clue. 
As shown in Fig.~\ref{fig:fig2}\textbf{a} and Movie~S1 for a separate experiment with higher temporal imaging resolution, as the colony undergoes fermentation, it nucleates a gas bubble at its upper surface (time $t=20$~h). This bubble grows, deforming the surrounding matrix and elongating vertically ($t=26$~h), until it eventually rises---entraining a vertical column of cells in its wake ($t=37$~h). Given that vertical dispersal only occurs during fermentation, not respiration, we hypothesize that the bubble is formed by CO$_2$ produced during fermentation. Fluorescence imaging via confocal microscopy confirms this hypothesis. Since dissolved CO$_2$ acidifies the surrounding matrix, we use sodium fluorescein, a pH-sensitive fluorophore [\emph{Supplementary Information}, Movie~S1], as a proxy for the local dissolved CO$_2$ concentration, $c_{\rm CO_2}$, focusing on a horizontal plane through the colony [dashed line in Fig.~\ref{fig:fig2}\textbf{a}]. As shown in Fig.~\ref{fig:fig2}\textbf{b}-\textbf{c}, the CO$_2$ produced by fermentation dissolves into the surrounding fluid, eventually reaching its saturation limit---triggering bubble nucleation and growth. A reaction–diffusion simulation that directly models dextrose fermentation and resulting CO$_{2}$ production by the colony corroborates these measurements, as shown in Fig.~S4\textbf{a}-\textbf{b} [\emph{Supplementary Information}].

\begin{figure}[h]
\centering
\includegraphics[width=0.95\textwidth]{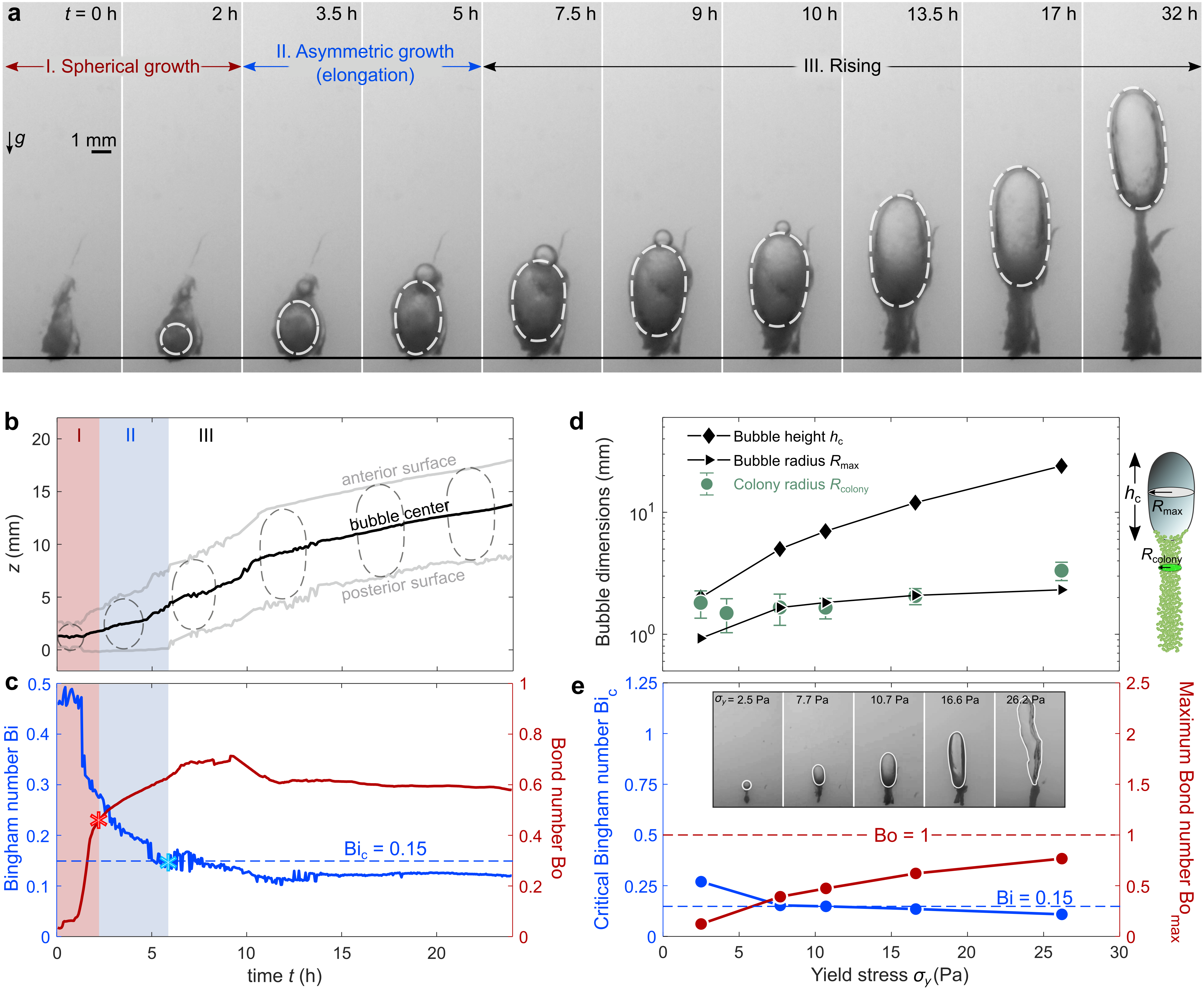}
\caption{\textbf{Capillary, buoyant, and yield stresses govern bubble shape, the onset of rising, and the radius of the dispersed colony.} \textbf{a,} Time series of bubble shape evolution in a matrix with $\sigma_y=10.7$~Pa ($c=1 \% ~\rm w/v$). Three distinct stages are observed: I. Spherical growth (red), governed by capillary stress; II. Asymmetric vertical elongation (blue), driven by buoyancy overcoming capillary stress; and III. Bubble rise following irreversible yielding of the surrounding matrix. \textbf{b,} Vertical trajectories of the bubble center, anterior surface, and posterior surface versus time for the experiment in \textbf{a}. \textbf{c,} Evolution of the Bond number $\text{Bo}$ and Bingham number $\text{Bi}$ during bubble growth and rise. The transition from stage I to II occurs when $\text{Bo}$ reaches $\text{Bo}_{\text{max}}\approx0.6$ (red asterisk); the transition from stage II to III occurs when $\text{Bi}$ drops to $\text{Bi}_c\approx0.15$ (blue asterisk, dashed line). \textbf{d,} Variation of bubble height $h_c$, bubble radius $R_{\text{max}}$, and colony radius $R_{\text{colony}}$ at the onset of rising with increasing matrix yield stress $\sigma_y$. Across all yield stresses, $R_{\text{colony}}\approx R_{\text{max}}$. \textbf{e,} Variation of $\text{Bo}_{\text{max}}$ and $\text{Bi}_c$ with $\sigma_y$; inset shows bubble shapes at the onset of rising for different $\sigma_y$, becoming increasingly irregular at higher yield stresses, reflecting a transition toward fracture-like cavity propagation. }\label{fig:fig3}
\end{figure}

What governs bubble shape and the onset of rising? And how do these processes influence the morphology of the final colony? 
As shown in Fig.~\ref{fig:fig3}\textbf{a–b} and Movie~S2c for an
experiment with even higher spatial and temporal imaging resolution, 
we observe three distinct stages of bubble dynamics as fermentation progresses. After it first nucleates, the bubble remains in place and grows spherically (stage I).
It then grows primarily vertically, elongating into an oblate ellipsoid while its bottom surface remains pinned at the colony (stage II). Finally, the elongated bubble detaches from the initial inoculum and rises through the matrix, entraining the colony in its wake (stage III).
To understand the transitions between these three stages, we compare the stresses that could govern bubble dynamics: the inertial stress $\sigma_i\sim\Delta\rho U_b^2$, viscous stress $\sigma_v\sim\mu_M U_b/R$, capillary stress $\sigma_c \sim \gamma/R$, buoyant stress $\sigma_b \sim \Delta \rho g h$, and yield stress $\sigma_y$, where $\Delta \rho\sim1~\mathrm{g}~\mathrm{cm}^{-3}$ is the density difference between the gas and surrounding matrix, $U_b\sim1~\upmu\mathrm{m}~\mathrm{s}^{-1}$ is the bubble rise speed, $\mu_M\sim10^4~\mathrm{Pa}\cdot\mathrm{s}$ is the effective matrix viscosity [\emph{Supplementary Information}], $\gamma\sim70~\mathrm{mN}~\mathrm{m}^{-1}$ is the fluid-gas surface tension, $R\sim1~\mathrm{mm}$ is the bubble radius,  $g$ is gravitational acceleration, and $h\sim1-10~\mathrm{mm}$ is the bubble height. Since the bubble rises slowly, the Reynolds number $\mathrm{Re} \equiv \Delta\rho U_b R/\mu_M\sim10^{-10}\ll1$, indicating that inertial stresses are negligible and the flow is in the creeping flow regime. Moreover, since the bubble is stationary during stages I and II, viscous stresses are negligible and the transitions between stages are governed purely by the interplay between capillary, buoyant, and yield stresses. Our experimental measurements of $R$ and $h$ enable us to directly track the evolution of these stresses.

In stage I (small $R\sim h$), $\sigma_c$ dominates, keeping the bubble spherical. As the bubble grows (increasing $R\sim h$), $\sigma_c$ decreases while $\sigma_b$ increases. Hence, their ratio, the Bond number $\mathrm{Bo}\equiv \frac{\Delta \rho g R^2}{\gamma}$, increases monotonically, as shown by the red curve in Fig.~\ref{fig:fig3}\textbf{c}. We expect that when $\mathrm{Bo}$ reaches a threshold value $\mathrm{Bo}_{\mathrm{max}}\sim \mathcal{O}(1)$, i.e., the buoyant stress becomes comparable to the capillary stress, the bubble can no longer maintain a spherical shape: buoyancy preferentially pushes the top of the bubble upward, driving vertical elongation. Our measurements [Fig.~\ref{fig:fig3}\textbf{c}] confirm this expectation. This transition to stage II, at which the bubble height continues to grow but its radius eventually saturates at $R_{\mathrm{max}}\sim\lambda_c =\sqrt{\gamma/\left(\Delta \rho g\right)}\approx2~$mm, where $\lambda_c$ is known as the capillary length, is indicated by the red asterisk in Fig.~\ref{fig:fig3}\textbf{c}. The corresponding Bond number plateaus at a maximum value $\mathrm{Bo}_{\mathrm{max}} \approx 0.6$.

As the bubble continues to elongate vertically in stage II (increasing $h>R_{\mathrm{max}}$), $\sigma_b$ continues to grow. Its ratio with $\sigma_y$, the Bingham number $\mathrm{Bi} \equiv \frac{\sigma_y}{\Delta \rho g h}$, decreases monotonically, as shown by the blue curve in Fig.~\ref{fig:fig3}\textbf{c}. For the bubble to escape confinement and rise, the buoyant stress must be sufficient to yield the surrounding matrix across the top surface of the bubble. Prior theoretical and numerical analyses established that this yielding must be sustained across an extended region surrounding the bubble; thus, $\sigma_b$ must exceed $\sigma_y$ by a factor of order $1/\mathrm{Bi}_{c}$, where $\mathrm{Bi}_{c}\sim \mathcal{O}(0.1)$ is the critical Bingham number for bubble rise in a yield-stress material~\cite{tsamopoulos2008steady, daneshi2023growth}. We therefore expect that when $\mathrm{Bi}$ drops below $\mathrm{Bi}_{c}$, the matrix can no longer confine the bubble, and begins to rise. Our measurements [Fig.~\ref{fig:fig3}\textbf{c}] also confirm this expectation. This transition to stage III, at which the bubble height eventually saturates at $h_{c} \sim 10\sigma_{y}/\left(\Delta\rho g\right)\approx10~$mm and it begins to rise, is indicated by the blue asterisk in Fig.~\ref{fig:fig3}\textbf{c}. The corresponding Bingham number falls to $\mathrm{Bi}_{c} \approx 0.15$.

These measurements indicate that the biogenic bubble shape and onset of rising are, to a first approximation, determined by the interplay between the capillary, buoyant, and yield stresses, as captured by the Bond and Bingham numbers. As a further test of this framework, we repeat this experiment across matrices of varying $\sigma_y$, measuring $R_{\mathrm{max}}$ and $h_c$ in each case [black triangles and diamonds, respectively, in Fig.~\ref{fig:fig3}\textbf{d}]. We use these measurements to determine $\mathrm{Bo}_{\mathrm{max}}$  and $\mathrm{Bi}_{c}$ [red and blue circles, respectively, in Fig.~\ref{fig:fig3}\textbf{e}]. Across different matrices, we expect the transitions between different stages of bubble dynamics to be governed by similar values of $\mathrm{Bo}_{\mathrm{max}}$  and $\mathrm{Bi}_{c}$ [\emph{Supplementary Information}]. Correspondingly, we expect $R_{\mathrm{max}}$ to be similar across all yield stresses, while by contrast, a larger yield stress would require a larger $h_c$ to trigger rising. We note, however, that our simple scaling analysis does not include the contribution of the matrix yield stress to resisting the deformation of the bubble from a sphere to an elongated cavity. At larger $\sigma_y$, a larger bubble is needed to generate sufficient buoyant stress to overcome this additional resistance, so we expect $\mathrm{Bo}_{\mathrm{max}}$ and $R_{\mathrm{max}}$ to increase slightly with $\sigma_y$---and the bubble shapes at the onset of rising to become increasingly irregular, reflecting a transition toward fracture-like cavity propagation~\cite{lee2020migration}. All our measurements [Fig.~\ref{fig:fig3}\textbf{d}-\textbf{e}, Movie~S2] agree well with these predictions, including the increasingly irregular bubble shapes at higher yield stresses [insets, Fig.~\ref{fig:fig3}\textbf{e}], confirming the validity of our physical framework.

\begin{figure}[h]
\centering
\includegraphics[width=0.9\textwidth]{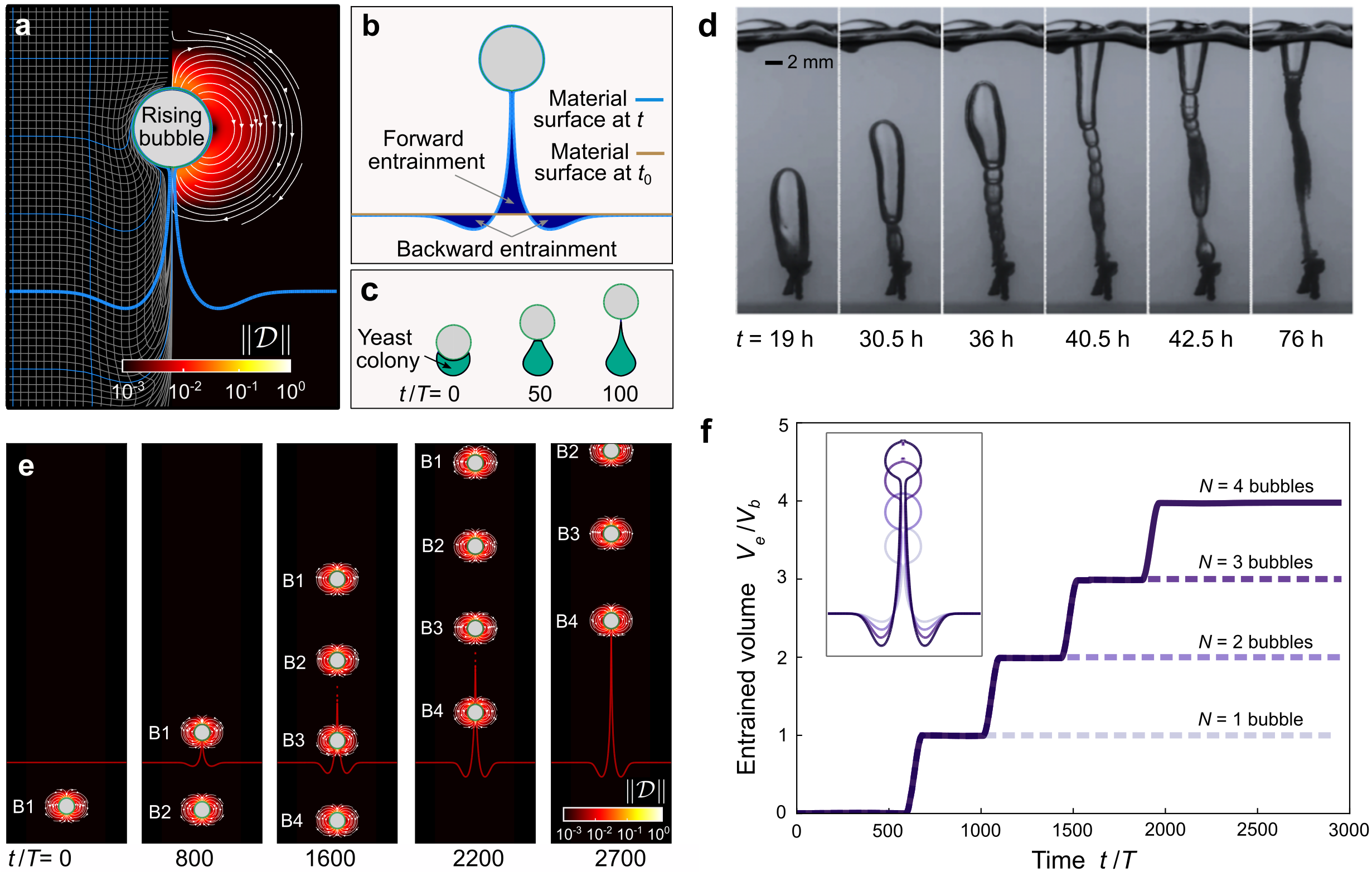}
\caption{\textbf{Sequential Darwin's drift by multiple bubble generations culminates in the formation of a columnar colony.} \textbf{a,} Simulation of fluid transport by a single bubble rising in a viscoplastic matrix ($\rm Bi=0.05$, $\rm Bo=1$). Left: deformation of an initially uniform rectangular grid of Lagrangian tracer particles after the bubble travels a vertical distance corresponding to 6.38 radii. Right: strain rate field $\left\|\bm{\mathcal{D}}\right\|$ overlaid with streamlines within the yielded zone. Unlike Stokes flow in a Newtonian fluid, deformation is confined to the yielded zone; the non-yielded region (dark) acts as a solid, enabling irreversible net displacement of surrounding material i.e., \textbf{b,} Darwin's drift in a viscoplastic medium. \textbf{c,} Lagrangian tracer simulation of colony entrainment by a single bubble, showing that a single bubble produces a conical colony profile, broader at the base and narrowing upward, inconsistent with the columnar morphology observed experimentally. The characteristic time scale is 
$T = \mu_M^0/(\Delta\rho g R_0)$ where $\mu_M^0$ is the matrix plastic viscosity and $\Delta\rho = \rho_M - \rho_B$ is the mass density difference between the matrix and the bubble.
\textbf{d,} Experimental time series of sequential bubble nucleation and rise in a matrix with $\sigma_{y}=16.6$~Pa ($c=1.2\% \rm w/v$). Secondary bubbles follow the weakened channel left by preceding bubbles. \textbf{e,} Simulation of sequential entrainment by four bubbles, showing that each additional bubble incrementally increases the total entrained volume and progressively transforms the conical profile into a columnar one. \textbf{f,} Cumulative drift volume as a function of time for $N=1-4$ bubbles; dashed lines show the drift volume contributed by each generation. Each successive bubble passage amplifies both forward and backward entrainment, collectively enabling sufficient dispersal to reach the air–matrix interface. }\label{fig:fig4}
\end{figure}

Beyond bubble shape, we also measure how the radius of the resulting vertically-dispersed colony varies in our experiments. We find that $R_{\rm colony}\approx R_{\rm max}$ (which in turn is $\approx\lambda_c$) across matrices of different yield stresses [green circles, Fig.~\ref{fig:fig3}\textbf{d}]. That is, unlike confined microbial colonies, whose shape is governed by surface-limited growth, here the colony radius is determined not by biology, but by physics---specifically, the competition between capillary and buoyant stresses on the biogenic bubble. 
This size matching suggests that the rising bubble entrains and transports cells upward via so-called \emph{Darwin's drift}, the net displacement of fluid caused by a moving body, which results in the  net transport of surrounding material~\cite{darwin1953note,eames1994drift,chisholm2017drift,jeanneret2016entrainment,mathijssen2018universal,zare2024bubble,katija2009viscosity}. In a Newtonian fluid at $\mathrm{Re}\ll1$, Darwin's drift extends well beyond the bubble, displacing fluid far from its path. Here, however, the nonlinear rheology of the viscoplastic matrix confines fluid displacement to the yielded zone surrounding the bubble; outside this zone, the matrix remains solid and cannot be displaced, restricting entrainment to a width that scales with the bubble radius and enabling the sharp, localized colony formation observed experimentally. Two-phase volume of fluid simulations, which incorporate both the nonlinear matrix rheology and the gas-liquid interfacial dynamics [\emph{Supplementary Information}], confirm this idea. As shown in Fig.~\ref{fig:fig4}\textbf{a}-\textbf{c} and Movie~S3, the rising bubble locally fluidizes the matrix, as indicated by elevated strain rate magnitudes $\left\|\bm{\mathcal{D}}\right\|$, while regions with $\left\|\bm{\mathcal{D}}\right\| \approx 0$ remain solid-like. This local yielding confines the fluid flow and enables both forward and backward displacement of fluid in a manner qualitatively distinct from Newtonian fluids at comparable Reynolds numbers [Fig. S7]. Thus, Darwin's drift by a biogenic bubble rising in a viscoplastic medium is sufficient to entrain cells vertically.

Lagrangian tracer particle tracks in the simulation show, however, that entrainment by a single bubble generates a \emph{conical} colony profile---broader at the base and narrowing upward [Fig.~\ref{fig:fig4}\textbf{c}, Movie~S3]. This shape is inconsistent with the columnar morphology of the colonies observed experimentally. A single bubble is therefore insufficient to account for the full extent of vertical colony dispersal. Indeed, as fermentation continues, the colony continues to produce successive bubbles [Fig.~\ref{fig:fig4}\textbf{d}, Movies~S2 and S4]. These secondary bubbles preferentially follow the same trajectory as the leading one, guided by the weakened channel of reduced yield stress left in its wake~\cite{lee2020migration}, forming a succession of rising bubbles that cumulatively increase the entrained volume. 
Simulations of multiple sequential bubbles confirm that each additional bubble contributes to the total drift volume [Fig.~\ref{fig:fig4}\textbf{e}–\textbf{f}, Movie~S5]. This cumulative effect allows for sufficient entrainment of the colony, enabling it to ultimately reach the air-matrix interface. This enhancement of entrainment is analogous to collective entrainment by schools of microswimmers, where sequential hydrodynamic interactions similarly amplify fluid transport~\cite{jin2021collective} and mixing~\cite{houghton2018vertically}.

\begin{figure}[h]
\centering
\includegraphics[width=0.9\textwidth]{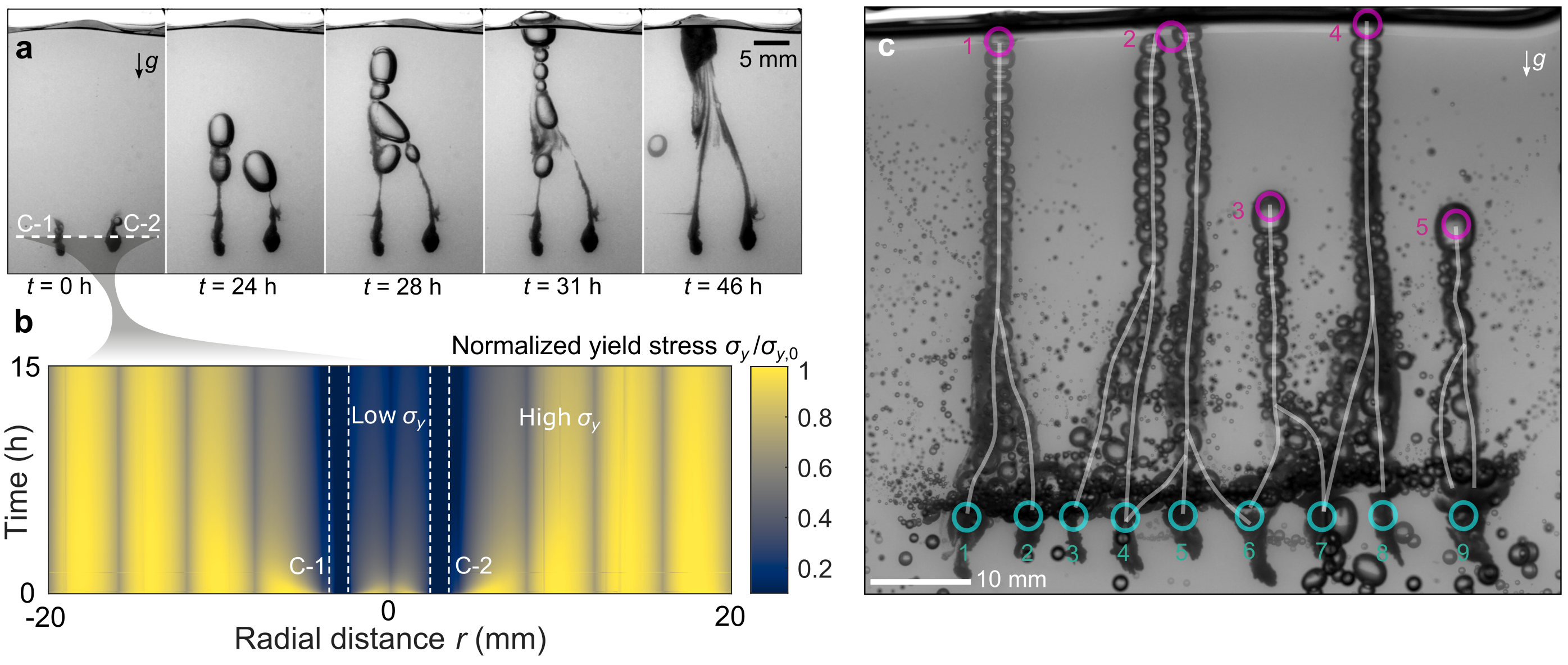}
\caption{\textbf{Intercolony interactions drive bubble steering, colony merging, and conduit network formation.} \textbf{a,} Time series of two yeast colonies positioned at the same height and separated by $\approx10~\mathrm{mm}$ (matrix $\sigma_{y}=7.7$~Pa, $c=0.9\% ~\rm w/v$). Bubbles from each colony deviate toward each other and eventually converge, causing the two colonies to coalesce. The dashed line indicates the horizontal plane at colony height where pH imaging was performed. \textbf{b,} Kymograph of local yield stress profiles normalized by the initial value, $\sigma_y/\sigma_{\rm y,0}$, measured via pH-sensitive confocal microscopy at the plane shown in
\textbf{a}. Fermentation produces a zone of lower pH, and thus lower yield stress, in the intercolony region, creating a mechanical gradient that steers bubbles toward the softer region. Periodic vertical stripes are imaging artifacts from digital stitching and do not reflect physical structure in the experiment. \textbf{c,} An array of nine colonies produces rising bubble streams that merge via yield-stress gradients between colonies, ultimately forming interconnected microbe-rich conduits that sustain continued biogenic gas release ($\sigma_y=7.7$~Pa, $c=0.9\% ~\rm w/v$).}\label{fig:fig5}
\end{figure}

Having established that a single colony achieves long-range vertical dispersal via entrainment by sequential biogenic bubbles, we next ask: Do multiple colonies interact within the same matrix, and if so, how? Many of the environments that confine microbes derive their mechanical integrity from charged polymer networks (e.g., EPS polysaccharides, humic substances) that can be weakened by fermentation byproducts~\cite{shakeel2022effect,costa2018microbial,adamczyk2009real,zhang2024microbial}. We confirm this effect in our hydrogel matrices, as well [Fig.~S3]. Therefore, we predict that when colonies are sufficiently close, their fermentation byproducts create a yield-stress gradient that steers their respective bubbles toward each other. We test this prediction directly by introducing two yeast colonies at the same height but separated by $\approx10~\mathrm{mm}$ horizontally. As shown in Fig.~\ref{fig:fig5}\textbf{a} and Movie~S6, bubbles generated by colonies 1 and 2 (C-1 and C-2, respectively) do indeed deviate toward each other, eventually converging---causing the two initially separate colonies to coalesce and mix genetically, in striking contrast to the genetic segregation typically observed in two-dimensional experiments~\cite{hallatschek2007genetic,korolev2010genetic}. We observe similar deviations even when a single bubble rises next to a colony [Fig.~S8, Movie~S4], indicating that this effect is not due to hydrodynamic interactions between bubbles~\cite{manga1995collective}; moreover, the strain rate associated with a bubble rising is sufficiently small that mechanical history effects in its wake can be ruled out~\cite{bhattacharjee2018polyelectrolyte}. Direct visualization confirms our prediction, showing a zone of lower pH between the colonies [dashed line in Fig.~\ref{fig:fig5}\textbf{a}, Movie~S6, Fig.~S3], as corroborated by reaction–diffusion simulations [Fig.~S4]. Converting the pH maps to yield stress profiles [\emph{Supplementary Information}] indicates time-dependent matrix softening near each colony [Fig.~\ref{fig:fig5}\textbf{b}], creating a mechanical gradient that steers bubbles toward the softer intercolony region~\cite{zare2021effects}. As a final test of this picture, we vary the initial inter-colony separation; as expected, the degree of bubble deviation increases systematically as colonies are placed closer together [Fig.~S9].

These bubble-mediated interactions can produce large-scale spatial structure. For example, nine colonies arranged in a linear array each produce rising streams of bubbles that merge due to the yield-stress gradients formed between the colonies---ultimately forming interconnected microbe-rich conduits [Fig.~\ref{fig:fig5}\textbf{c}, Movie~S7]. These conduits are self-sustaining: continued metabolic activity generates new bubbles in or near existing conduits, reinforcing the conduit structure as the bubbles rise along this path of least resistance. Intriguingly, analogous conduit networks have been observed across a range of natural systems, but the biophysical mechanisms underlying their formation have remained unclear. In fermenting bread dough, X-ray tomography has revealed that yeast-generated CO$_{2}$ bubbles coalesce into a single interconnected cluster that drives subsequent dough expansion~\cite{babin2006fast,chakrabarti2021flour}---yet the potential role of metabolically driven yield-stress gradients in steering this coalescence, and in generating microbe-rich conduits, has not been recognized. In aquatic sediments, methane ebullition is notoriously clustered at persistent spatial hotspots~\cite{scandella2011conduit,wik2016climate,davidson2018synergy}, releasing both methane and microbes into the water column~\cite{schmale2015bubble,jordan2020bubble}---yet the mechanisms underlying both spatial clustering and entrainment have eluded explanation. Our findings raise the hypothesis that in both cases, metabolic activity generates bubbles that entrain and transport microbial cells through the viscoplastic matrix, while the spatial heterogeneity in matrix yield stress generated by continued metabolic activity steers subsequent bubbles along the same paths, shaping and reinforcing the conduit networks observed in these systems. Our work may also help explain the finger-like, columnar morphologies long observed in microbial mats~\cite{bosak2010formation,bosak2009morphological,juarez2025morphology,voorhies2012cyanobacterial}, where biogenic bubbles vertically entrain an entangled network of microbes by an analogous process. Investigating whether the mechanisms identified here operate similarly in natural settings---where matrix composition, metabolic diversity, and rheology are more complex than in our model experiments, and where the direction of pH-driven yield stress changes may differ---will be an important direction for future research. 

Our work opens several other directions for future investigation. At the single-colony level, our experiments revealed the pivotal role of Darwin's drift in entraining cells through the viscoplastic matrix; characterizing how the efficiency of this entrainment depends on matrix rheology across the full parameter space, and extending our theoretical framework to 3D, will be a useful next step. At the collective level, it remains to be understood how conduit network topology evolves as the number of colonies increases, and whether isolated conduits merge into a system-spanning network above some critical colony number density. From an ecological perspective, the finding that the radius of the dispersed colony is governed by the physical length scale $\lambda_c$, rather than by biology, raises the question of how spatial variations in matrix properties shape the probability of colony merging, genetic mixing, and competition outcomes in natural settings.

Metabolic activity is a defining feature of life. And yet, its capacity to act as a driver of active matter behavior through mechanochemical coupling to the surrounding environment is only just beginning to be recognized. The canonical classification of active matter---collectives of agents that actively consume energy to exert mechanical forces on their environment---focuses on two mechanisms of force generation: motility~\cite{marchetti2013hydrodynamics,bechinger2016active} and growth~\cite{hallatschek2023proliferating}. Our findings highlight a fundamentally distinct mode of active behavior that does not fit naturally within this canonical classification, joining a growing body of observations in which metabolic byproducts drive flows and dispersal across diverse systems~\cite{david2025explosive,narayanasamy2025metabolically,atis2019microbial,chen2024collective,fragkopoulos2025metabolic,yan2017extracellular}. Taken together, these observations motivate the introduction of a third class: \emph{Metabolically Driven Active Matter}, in which mechanical forces are generated not by motility or growth, but by metabolic byproducts that alter the physical properties of the surrounding environment, allowing individual cell metabolism to drive colony-scale motion. Our work reveals that the chemical products of microbial metabolism can reshape the physical landscape of confinement itself---converting an impenetrable barrier into a network of conduits that sustain dispersal, transport, and community structure across environments as diverse as soils, sediments, and fermented foods~\cite{jerolmack2019viewing,newman2002geomicrobiology,falkowski2008microbial,singh2010microorganisms,reay2018methane,krevor2023subsurface,tyne2021rapid}.

\backmatter






\bmhead{Supplementary Movies} 
The supplementary movies are available at \url{https://zenodo.org/records/19393890}. 


\bmhead{Acknowledgments} 
We acknowledge support from National Science Foundation (NSF) grants CBET-1941716, DMR-2011750, and EF-2124863 as well as the Camille Dreyfus Teacher-Scholar and Pew Biomedical Scholars Programs. We thank R. K\={o}nane Bay for assistance with preliminary experiments, Marjan Zare and Ian Frigaard for assistance with preliminary simulations, as well as Hugo Leonardo Fran\c{c}a, Chris MacMinn, Arnold Mathijssen, Howard Stone, and members of the Datta Lab for stimulating discussions and useful feedback.

\bmhead{Author Contributions} 
B.V.H. and S.S.D. conceptualized and designed the overall research project; B.V.H., H.N.L., and M.R. performed all experiments and experimental analyses; T.A. and M.J. performed the numerical simulations; B.V.H., T.A., M.S., M.J., and S.S.D. wrote the article.

\bmhead{Competing Interests} 
The authors declare no competing interests.

\clearpage
\setcounter{figure}{0}
\setcounter{table}{0}
\renewcommand{\thefigure}{S\arabic{figure}}
\renewcommand{\thetable}{S\arabic{table}}

\begin{center}
{\Large\bfseries Supplementary Information}
\end{center}

\section{Materials and methods}\label{sec1}

\subsection{Cell culturing and sample preparation}
We use wild-type \textit{Saccharomyces cerevisiae} strain CEN.PK2-1C throughout. Cells are cultured overnight in standard YPD growth medium consisting of 1\% yeast extract, 2\% peptone, and 2\% dextrose (unless otherwise stated). Following overnight growth, the cultures are centrifuged at 3000 rpm for 1 minute, and the supernatant is removed. A 5 µL aliquot of the cell pellet is then embedded into the hydrogel matrix.
The granular hydrogel matrix is prepared using Carbopol® 980, a crosslinked poly(acrylic acid/alkyl acrylate) copolymer. In its dry form, Carbopol consists of collapsed, internally-crosslinked polymer granules that swell upon dispersion in aqueous media. After mixing with liquid growth media, the carboxylic acid groups in the polymer chains are neutralized using 10 M NaOH, promoting electrostatic repulsion that expands the microgel network and enhances interparticle interactions. The swelling of the hydrogel particles (diameter $\approx$ 5–10 µm) renders the matrix optically transparent. The final mechanical properties---pore size and yield stress---are set by the polymer concentration, with higher Carbopol concentrations yielding more jammed matrices with smaller pores and larger yield stresses.
Carbopol powder is dispersed directly into YPD medium at concentrations ranging from 0.7–1.5 wt\%. The mixture is stirred for at least 24 hours to ensure homogeneity. To adjust the pH back to $\approx$5, the native pH of yeast growth, a volume of 10 M NaOH (42–65 µL per 10 mL of dispersion) is added, depending on Carbopol concentration.

For sample preparation, we fill a 25 mL tissue culture flask with the hydrogel suspension up to a height of $\approx$ 10 mm using a syringe with a 12-gauge needle. Then, 5 µL of the centrifuged yeast pellet is injected $\approx$ 5 mm below the surface of the prepared hydrogel matrix using a pipette that is held vertically during injection and slowly withdrawn afterward. Then, additional hydrogel suspension is gently layered on top until the desired volume was reached. All imaging experiments are performed at room temperature. This preparation method avoids causing mechanical damage to the matrix that could potentially alter the bubble migration trajectory.

To image spatial variations in pH, dissolved CO$_2$ concentration ($c_{\rm CO_2}$), and local yield stress ($\sigma_{y}$), hydrogel matrices are deposited into glass-bottom dishes (Cellvis). Then, using a pipette, we inject one or two vertically oriented columnar yeast colonies. The injected colonies are cylindrical in shape, with thicknesses comparable to those observed in bubble-driven colony morphologies in the main experiments.
For these measurements, the initial pH of the growth medium is adjusted to 6.5, higher than in other experiments (pH$\sim5$), to better match the sensitivity range of the pH-responsive fluorescent dye, sodium fluorescein. 

\subsection{Rheology of the granular hydrogel matrix}
\begin{figure}[h]
        \includegraphics[width=1\textwidth]{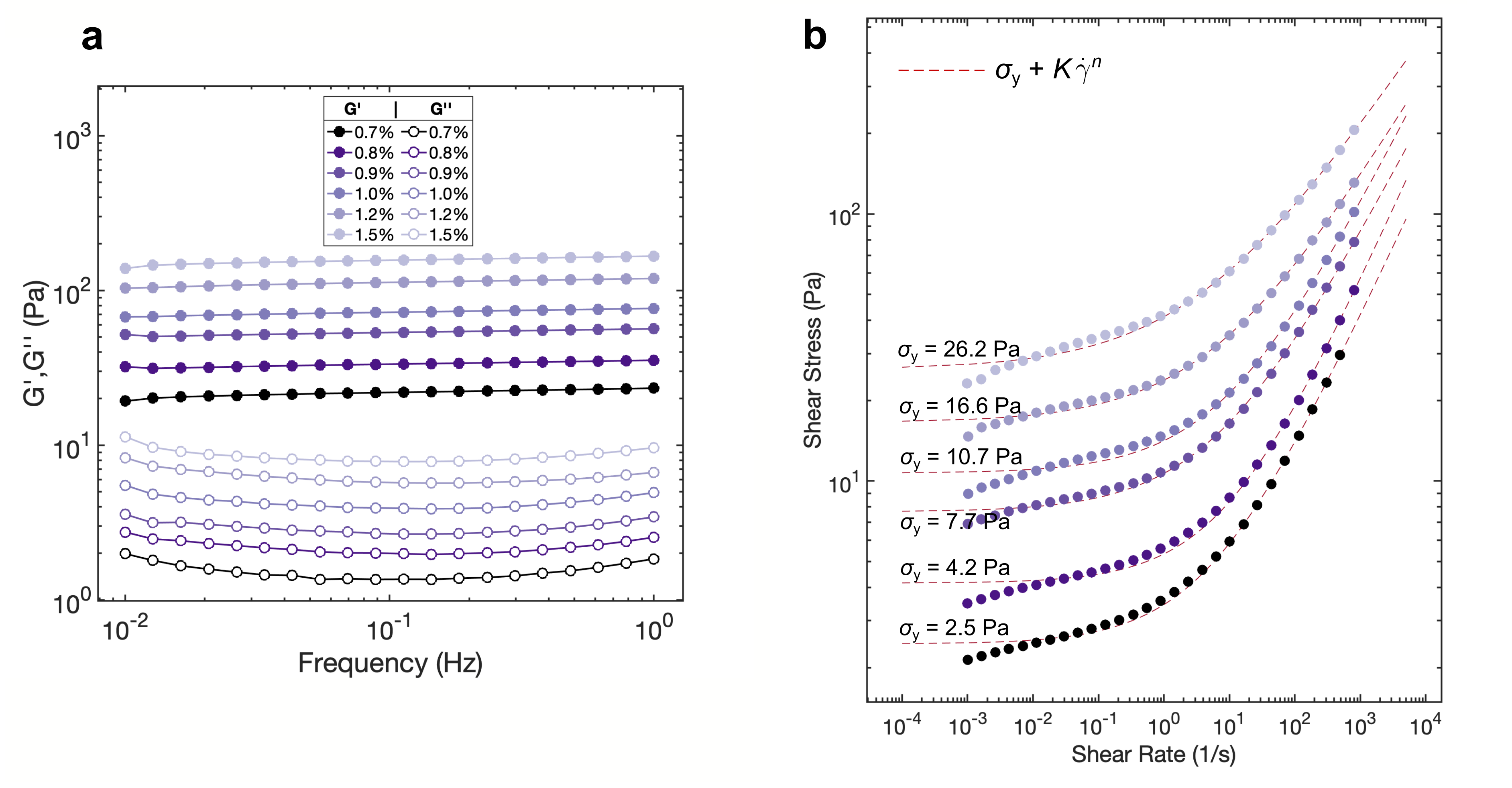}
        \caption{
            Rheology of hydrogel matrices. A) The storage ($G'$) and loss ($G''$) moduli of each Carbopol concentration (in $\rm w/v\%$) against frequency at low strain. Storage moduli are higher than loss moduli at all frequencies measured, indicating that the material behaves as an elastic solid at low strain. B) Shear stress of each Carbopol concentration plotted against unidirectional shear rates. The curves are fitted with the Herschel-Bulkley model for shear-thinning, viscoplastic fluids, where $\sigma_{y}$ represents the yield stress, \textit{K} is the consistency index, \textit{n} is the flow index, and $\dot{\gamma}=\mathcal{D}$ is the shear rate. The specific values of each variable obtained from the fit are presented in Table~\ref{tab:Rheology1}. 
        }
        \label{fig:Rheology1}
    \end{figure}

To characterize the hydrogel matrices, we measure the storage and loss moduli as well as the flow curves using an Anton Paar MCR 501 rheometer.  We load approximately 2 - 3 mL of a given hydrogel matrix into the 1 mm gap between 50 mm-diameter
parallel plates. To minimize wall slip, we use a roughened top plate and attach sandpaper to the bottom measuring plate using double sided tape. We perform oscillatory shear measurements by sweeping the frequency from 0.1 - 1Hz at a fixed strain of 1\%  to determine the storage modulus, $G'$, and the loss moduli, $G''$ (Fig.~S\ref{fig:Rheology1}\textbf{a}). For all the matrices used in the experiments, the storage modulus exceeds the loss modulus $G''$ by an order of magnitude, indicating that the material is a jammed solid at zero shear. We expect these materials to fluidized at higher shear rates, and to characterize this behaviour, we measure the shear stress at different shear rates from $10^{-3} - 10{^3} \ \text{s}^{-1}$ as shown in Fig.~S\ref{fig:Rheology1}\textbf{b}. We find that at low shear rates, the shear stress remains nearly constant and is independent of the shear rate, indicating a finite yield stress -- this sets the force scale for bubble motion. As we increase the shear rate, the  matrices fluidize, and the shear stress follows a power law dependence on the shear rate.  We can fully parameterize this flow curve using the classic Herchel-Buckley equation (red dashed lines). Using this approach, we generate transparent viscoplastic media with tunable yield stresses ranging from $2-30$ Pa. 

\begin{table}
\begin{tabular}{p{4cm}p{3cm}p{3cm}p{1cm}}
 \hline 
 \hline
Carbopol Conc. ($\rm w/v\%$) & $\sigma_{y} \ (\rm Pa)$ & $K \ (\text{Pa}\ s^n)$ & $n$ \\
 \hline
 0.7 & 2.45 & 0.99 & 0.53 \\
0.8 & 4.15 & 1.19 & 0.55  \\
0.9 & 7.67 & 3.01 & 0.47  \\
1.0 & 10.7 & 3.46 & 0.49  \\
1.2 & 16.58 & 7.27 & 0.41  \\
1.5 & 26.16 & 15.6 & 0.37  \\

 \hline

 \end{tabular}
 \caption{Values of yield stress, consistency index, and flow index obtained from Herschel-Bulkley model fit.}
    
    \label{tab:Rheology1}
\end{table}

\subsection{Imaging}
The optical transparency of the granular hydrogel matrix allows direct visualization of both colony growth dynamics and biogenic bubble formation. To capture the long-timescale dynamics at the centimeter scale, we use three parallel imaging setups for simultaneous data acquisition.
Each setup consists of a Nikon Micro-NIKKOR 55mm f/2.8 lens mounted on a Sony $\alpha 6300$ camera, with illumination provided by a uniform LED light panel. Time-lapse imaging is performed at 60-second intervals over a period of approximately 100 hours, generating multicolor image sequences.
Two illumination configurations are used. For bright-field imaging (e.g., Fig. 1), the LED panel is positioned on the front side of the sample, producing high-contrast images of the colony against a dark background. For shadowgraphic imaging (used in Figs.1–5), the LED panel is placed behind the sample, resulting in bright-field images where the colony and the bubbles appear dark in a brighter background.
The acquired images provide two-dimensional projections of the growing yeast colony and rising bubbles. Image sequences are binarized and analyzed using ImageJ and MATLAB to extract morphological metrics including colony area, mean colony diameter, and bubble dimensions.

To visualize the spatial distribution of dissolved carbon dioxide $c_{\rm CO_2}$, (Fig. 2\textbf{b}) and infer local yield stress variations in the hydrogel matrix (Fig. 5), we use a Nikon AXR inverted laser scanning confocal microscope and a pH-sensitive fluorescent dye, sodium fluorescein.
Imaging is performed using a $4\times$ objective lens at a rate of 1 frame per minute.  Fluorescent images are acquired from a horizontal optical slice (depicted as dashed lines in the corresponding figures) of approximately $100~ \upmu \rm m$ thickness. To capture the full spatial extent of CO$_2$ diffusion within the matrix, large-area scanning is performed via automated frame stitching. The vertical stripes visible in Fig. 5\textbf{b} correspond to frame boundaries introduced during the image stitching process.

\subsection{pH measurement and calibration to obtain $c_{\rm CO_2}$}

\subsubsection{Calibration and conversion of fluorescence intensity to pH}
\begin{figure}[h]
\centering
        \includegraphics[width=0.4\textwidth]{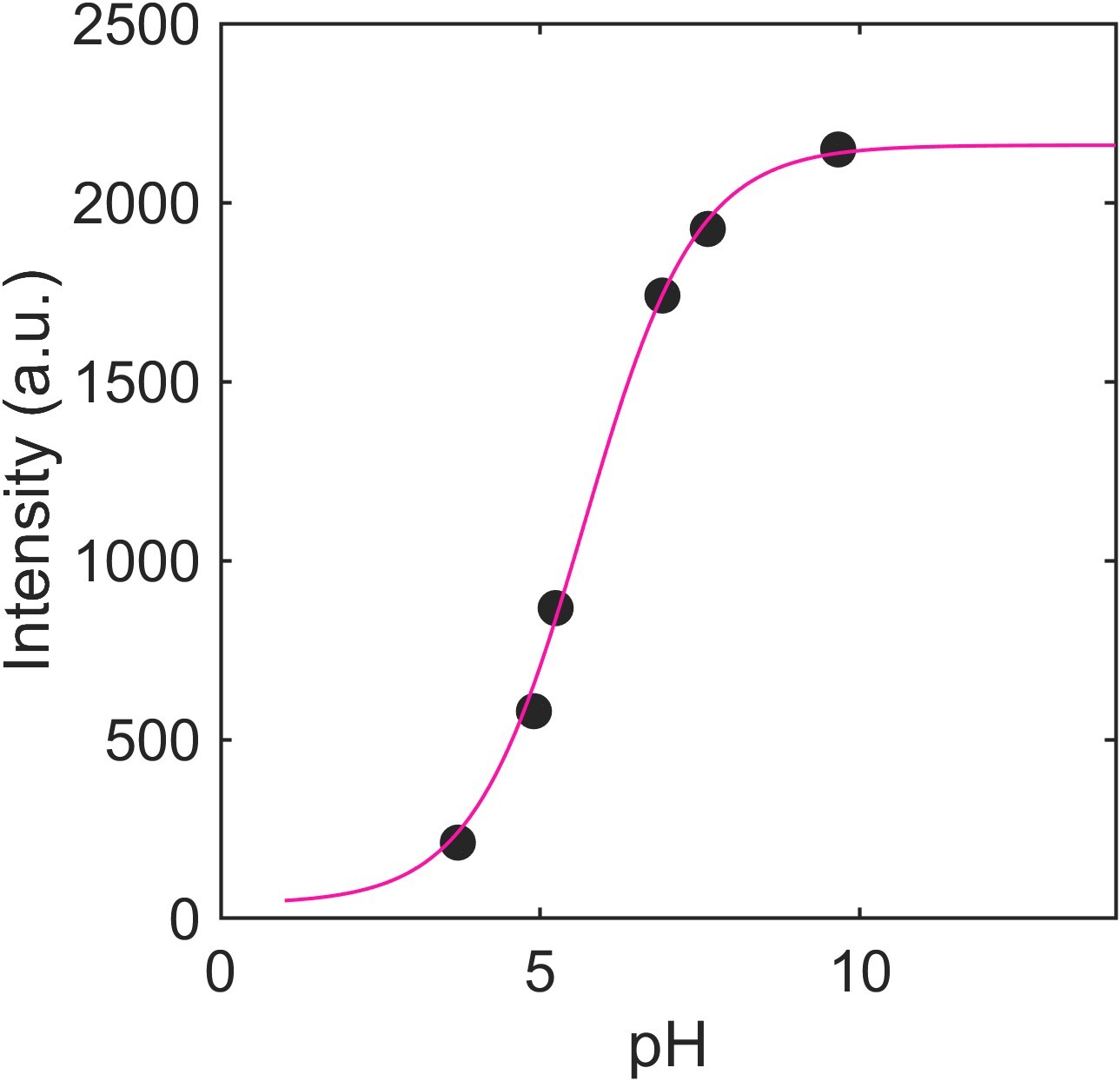}
        \caption{
            Calibration and conversion of fluorescence intensity to pH. Data points represent the mean measured fluorescence signal at each pH value. The magenta curve shows the best-fit sigmoid function to the experimental data. 
        }
        \label{fig:Calibration_CO2}
    \end{figure}

To calibrate the fluorescent images acquired by confocal microscopy, we prepare seven hydrogel matrix samples in YPD medium at pH values of 3.71, 4.90, 5.24, 6.04, 6.91, 7.62, and 9.66, each containing the same concentration of sodium fluorescein dye. pH values are measured using a Fisherbrand Accumet AE150 pH meter. Fluorescent images of each sample are obtained using the same imaging parameters as in Figs.~2 and~5, as described in the \textit{Imaging} section. The mean image intensity is shown against pH in Fig.~S\ref{fig:Calibration_CO2}. A calibration curve converting local fluorescence intensity to pH is obtained by fitting the data to a sigmoid function,
\begin{equation}
\label{eq:Int_to_pH}
I = \frac{3.18}{\exp(-1.1441\,pH) + 0.0015} + 40 .
\end{equation}
The inverse relation,
\begin{equation}
\label{eq:pH_to_Int}
pH = -\frac{1}{1.1441} \ln\left(\frac{3.18}{I - 40} - 0.0015\right),
\end{equation}
then enables us to calculate local $c_{\rm CO_2}$ (Fig.~2, detailed in \textit{Supp. Info.}~1.4.2) or local $\sigma_{y}$ (Fig.~5, detailed in \textit{Supp. Info.}~1.5).

\subsubsection{Calculation of the pH conversion to $c_{\rm CO_2}$}
To convert the change in the pH of the media to the concentration of $c_{\rm CO_2}$, we consider the equilibrium reactions that describe the dissolution of \ce{CO2}. The \ce{CO2} produced by the yeast dissolves to produce carbonic acid, which then dissociates to form bicarbonate and \ce{H^+} ions:
\smallskip
\begin{center}
    \ce{H2O + CO2 <=> H2CO3 <=>  H^+ + HCO3^-}.
\end{center}
The bicarbonate further dissociates to form carbonate:
\smallskip
\begin{center}
    \ce{HCO3^- <=> H^+ + CO3^2-}.
\end{center}
\smallskip
The total \ce{CO2} produced by the yeast is the sum of all the dissolved inorganic carbon in the system, $c_{\rm CO_2}$ = \ce{[CO2] + [H2CO3] + [HCO3^-] + [CO3^2-]}. To estimate this concentration, we consider the equilibrium kinetics of the system~\cite{averill2012principles}. Due to the rapid dissociation of carbonic acid, we neglect the undissociated, neutral aqueous carbonic acid~\cite{adamczyk2009real, stefansson2013carbonic, wang2016stable, johnson1992supcrt92}, and instead consider the reaction 
\smallskip
\begin{center}  
    \ce{H2O + CO2 <=>  H^+ + HCO3^-}.\\
\end{center}
\smallskip
with an effective dissociation constant, $K_1 = 4.5 \times 10^{-7}$, which governs the production of bicarbonate
\begin{equation}
\label{eq:K1Equation}
    K_1 = \frac{\ce{[H^+] [HCO3^-]}}{[\ce{CO2]}}
\end{equation}
The second dissociation constant, $K_2 = 4.7 \times 10^{-11}$, governs the production of carbonate:
\begin{equation}
\label{eq:K2Equation}
    K_2 = \frac{\ce{[H^+] [CO3^2-]}}{[\ce{HCO3^-]}}
\end{equation}
We can solve Eq. ~\ref{eq:K1Equation} and Eq.~\ref{eq:K2Equation} to determine the concentration of bicarbonate and carbonate ions in the solution. 
\begin{equation}
\label{eq:BicarbConc}
    \ce{[HCO3^-]} = \frac{K_1\ce{[CO2]}}{\ce{[H^+]}}
\end{equation}
and
\begin{equation}
\label{eq:CarbConc}
    \ce{[CO3^2-]} = \frac{K_2\ce{[HCO3^-]}}{\ce{[H^+]}} = \frac{K_1K_2\ce{[CO2]}}{\ce{[H^+]}^2}
\end{equation}
\smallskip
Assuming the system is not buffered, the production of \ce{CO3^2-} and \ce{HCO3^-} will result in the production of an equal amount of \ce{H^+} ions that is responsible for altering the pH of the system. Thus, 
\begin{equation}
\label{eq:TotHConc}
    \ce{[HCO3^-]} + \ce{[CO3^2-]} = \ce{[H^+]}_{final} - \ce{[H^+]}_{initial} = 10^{-\rm pH} - 10^{-\rm pH_0}  
\end{equation}
where $\rm pH_0$ is the initial pH of the system. We note that YPD medium contains peptone and yeast extract, which have weak buffering capacity; we neglect this buffering here as a first approximation, which may lead to a modest underestimate of $c_{\rm CO_2}$. Using Eq.~\ref{eq:BicarbConc}, ~\ref{eq:CarbConc}, and ~\ref{eq:TotHConc}, we can solve for the total carbon dioxide produced by the colony
\begin{equation}
    \label{eq:TotConc}
    c_{\rm CO_2} = \left( 10^{-\rm pH} - 10^{-\rm pH_0}   \right)\left( \frac{10^{- \rm pH}}{K_1}\left(\frac{1}{1 + K_2/10^{- \rm pH}}\right) + 1\right) \approx \left( 10^{-\rm pH} - 10^{-\rm pH_0}   \right)\left( \frac{10^{-\rm pH}}{K_1} + 1\right)
\end{equation}
as $K_2 \ll 10^{-\rm pH}$ for our system.

\subsection{pH measurement and calibration for local yield stress}

\begin{figure}[h]
\centering
        \includegraphics[width=0.9\textwidth]{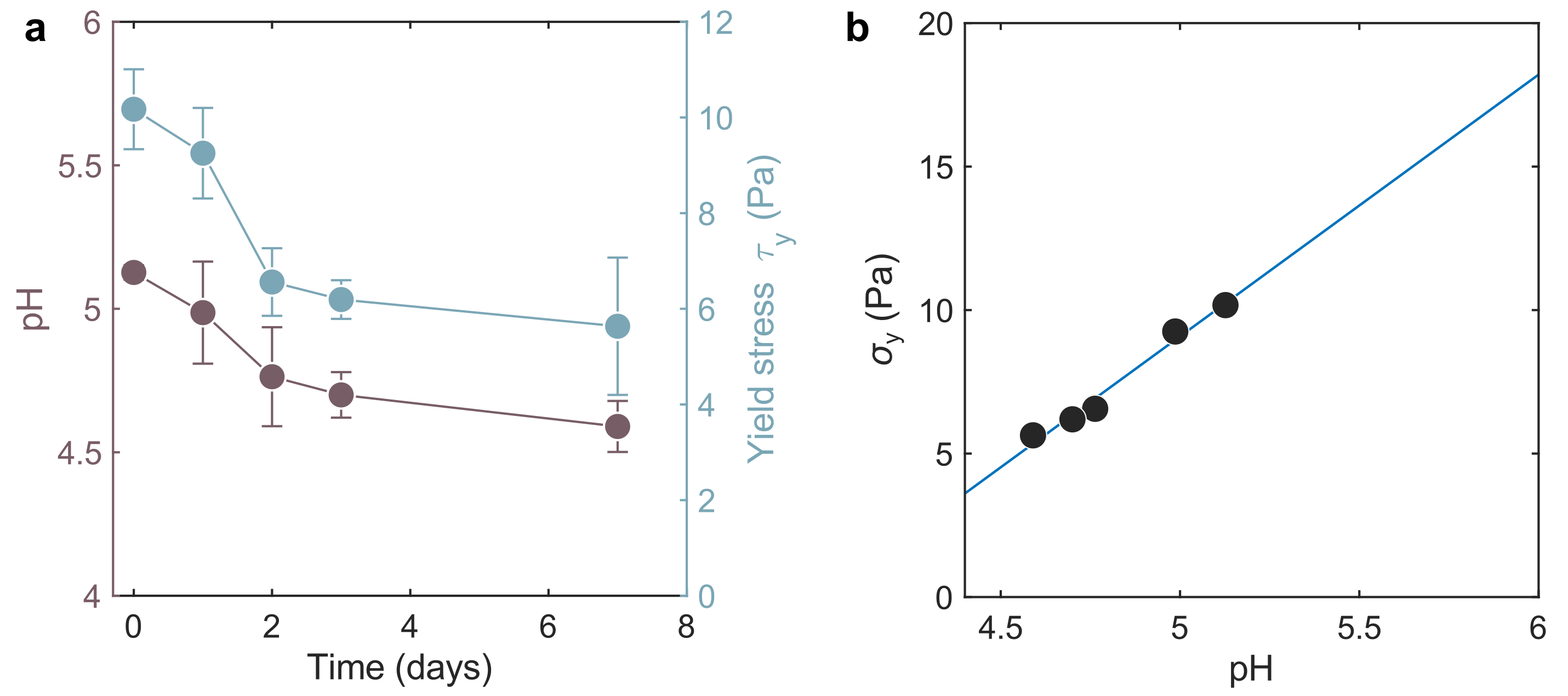}
        \caption{
            Calibration and conversion of fluorescence intensity to local yield stress. \textbf{a,} Experimental measurement of the variation of pH and yield stress of the hydrogel matrix embedded with uniformly distributed yeast colonies. Measurements were conducted 0, 1, 2, 3, and 7 days after the onset of fermentation.  The error bars represent the standard deviation of 3 replicates. \textbf{b,} Calibration curve, a linear fit $\sigma_{y}=  9.1226~pH - 36.53$, using values from experiments in panel \textbf{a}, is used to convert local pH to local yield stress values. 
        }
        \label{fig:Calibration_sigma}
    \end{figure}

To test whether the metabolic activity by the yeast modifies the mechanical properties of the matrix, we examine how yeast metabolism affects the yield stress of the hydrogel matrices as well as their pH values. We prepare 15 samples of hydrogel matrices with $c=1 \% ~\rm w/v$ and uniformly disperse $5 \upmu \text{L}$ of the yeast pellet in the matrix.
We let the yeast perform fermentation for 0, 1, 2, 3, and 7 days (3 samples for each fermentation period) and then measure the pH and the yield stress of the matrix.
The fermentation byproducts CO$_2$ and ethanol acidify the medium over time, leading to a reduction in yield stress (Fig.~S\ref{fig:Calibration_sigma}\textbf{a}). This measurement shows that fermentation tends to soften the hydrogel matrix. Furthermore, it can serve as a calibration to convert the local pH values to the local yield stress values within the matrix.
As shown in Fig.~S\ref{fig:Calibration_sigma}\textbf{b}, we obtain the calibration curve $\sigma_{y}=  9.1226~pH - 36.53$. Plugging the pH from Eqn. \ref{eq:pH_to_Int} into this calibration curve yields the calibration relation to convert local image intensity to local yield stress in Movie S6:
\begin{equation}
\label{eq:Int_to_sigma}
\sigma_{y}=-6.33 \ln \left( \frac{3.18}{I}-0.0015\right) -36.53.
\end{equation}

\section{Theory and Simulations}\label{sec2}

\subsection{Estimates for surface-limited growth}

To estimate the biomass production by a yeast colony suspended in the hydrogel matrix, we assume a spherical colony $R_{\rm col} = 1~ \rm mm$ that is densely packed with yeast cells. At such a cell concentration, growth is limited to a thin outer layer of the colony. We assume a constant doubling time within this layer and use measured values obtained from the literature~\cite{olivares2018saccharomyces} for dextrose and glycerol. In such conditions, the radius grows linearly with time \textit{t}:
\begin{equation}
R_{\rm col}(t) =R_{\rm col,0} + k_{\rm g}\ l_{\rm n}\ t
\end{equation}
where $R_{\rm col,0}$ is the initial radius of the colony, $k_{\rm g}$ is the growth rate, and $l_{\rm n}$ is the nutrient penetration depth as calculated in the main text. We can now estimate the biomass of the colony using the expression above:
\begin{equation}
\text{Biomass(\textit{t})} = \frac{4}{3} \pi ~(R_{\rm col,0} + k_{\rm g}\ l_{\rm n}\ t)^3
\label{eqn.:biomass}
\end{equation}

We choose estimated values of $R_{\rm col,0}= 1.5$ mm, $l_{\rm n}=20 \rm~\upmu m$, and $k_{\rm g}= 0.15 ~\rm hr^{-1}$ for growth in YPD medium (dextrose as nutrient, which favors fermentation) and $k_{\rm g}= 0.059 ~\rm hr^{-1}$ for growth in YPG medium (glycerol as nutrient, which favors respiration). Eq.~(\ref{eqn.:biomass}) yields $\frac{\rm Biomass}{\rm Biomass_{\textit{t}=0}}\approx 3.24$ in YPD medium and $\frac{\rm Biomass}{\rm Biomass_{\textit{t}=0}}\approx 1.68$ in YPG medium.

\subsection{Reaction-diffusion simulations for the metabolic activity and diffusion of the dissolved CO$_2$}
We estimate the increase in CO$_2$ concentration resulting from fermentation using a simple 1D spherical continuum model that accounts for glucose diffusion and consumption, as well as the coupled production of  CO$_2$. We assume that the cells are close packed and localized at $r = 0$,  with a radius of $1$ mm.  The nutrient distribution mirrors experimental conditions: the region containing cells is initially depleted of glucose, while the surrounding region maintains a uniform concentration , $c_{\rm g, 0}$. Cells consume glucose at a maximum rate $\kappa$ modulated by local nutrient availability through Michaelis–Menten kinetics:
\begin{equation}
        \frac{\partial c_{\rm g}}{\partial t} = D_{\rm g} \frac{1}{r^2}\frac{\partial}{\partial r}\left(r^2\frac{\partial c_{\rm g}}{\partial r} \right) - \kappa \rho  \frac{c_{\rm g}}{\left( c_{\rm g}^* + c_{\rm g} \right)}
\end{equation}
where $r$ is the radial position,  $D_{\rm g}$ is the diffusion coefficient of glucose, and $c_{\rm g}^*$ is the Michaelis-Menten half-saturation constant. 

The concentration of CO$_2$, $c_{\rm CO_2}$, is governed by the production by the cells based on local glucose availability and diffuses away from the colony center: 
\begin{equation}
        \frac{\partial c_{\rm CO_2}}{\partial t} = D_{\rm CO_2} \frac{1}{r^2}\frac{\partial}{\partial r}\left(r^2\frac{\partial c_{\rm CO_2}}{\partial r} \right) +\alpha \rho  \frac{c_{\rm g}}{\left( c_{\rm g}^* + c_{\rm g} \right)}
\end{equation}
where $\alpha$ is the production rate of glucose and $D_{\rm CO_2}$ is the diffusion of carbon dioxide. We impose radial symmetry at $r = 0$ and apply no-flux boundary conditions for both glucose and CO$_2$ at the outer edge of the domain. We run simulations for 24 hours using the parameter values listed in Table~\ref{tab:PenLength}.

\begin{figure}[h]
        \includegraphics[width=1\textwidth]{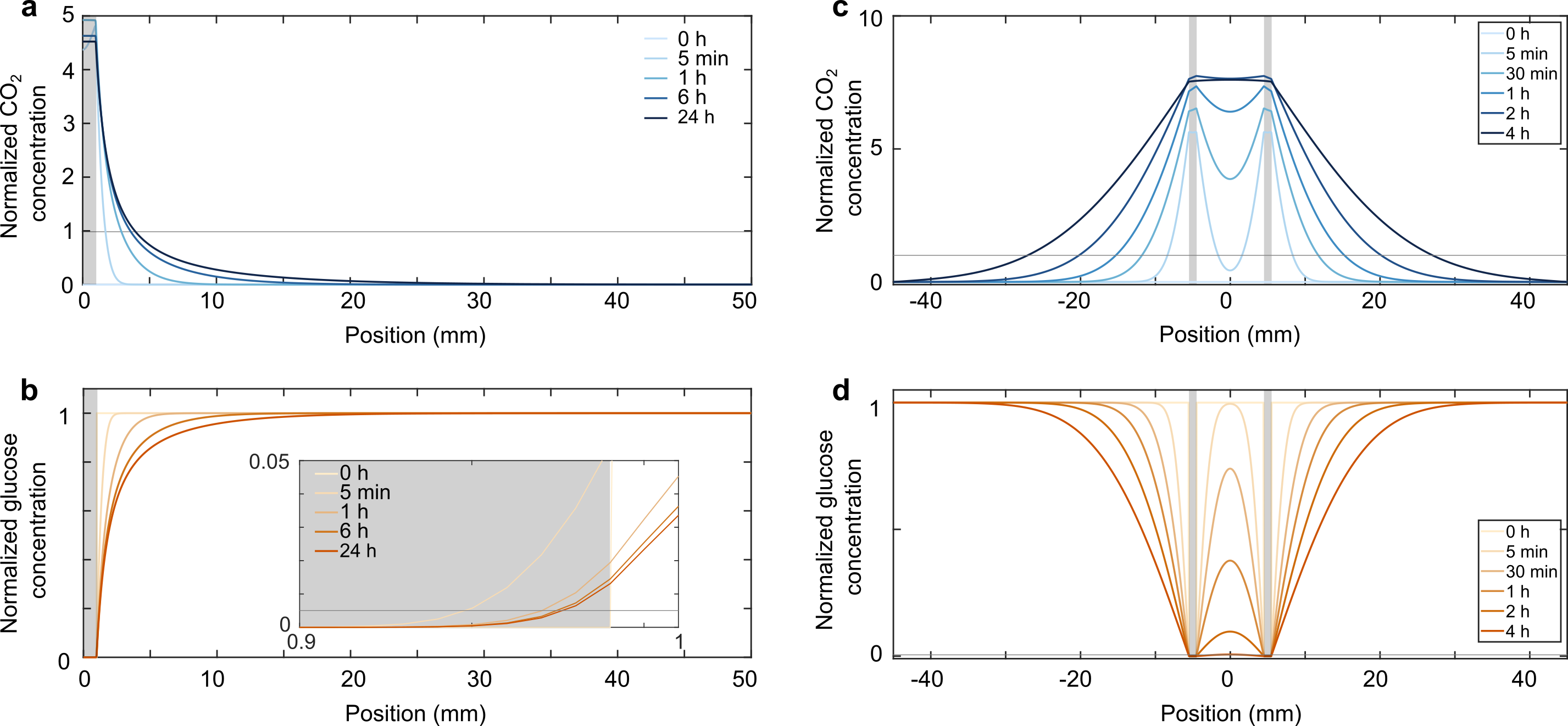}
        \caption{
            Simulations of the metabolic activity by the colony and the diffusion of the metabolites in the environment. The carbon dioxide and glucose profiles at different time points are plotted for a single colony in panels \textbf{a} and \textbf{b}, and for two neighboring colonies in panels \textbf{c} and \textbf{d}. The shaded region indicates the location of the colonies. The inset in panel \textbf{b} shows the zoomed-in view of the concentration profiles near and within the colony. The carbon dioxide concentration is normalized by the saturation concentration (gray horizontal lines in panels \textbf{a} and \textbf{c}), and the glucose concentration is normalized by the initial value. The solid gray lines in panels \textbf{b} and \textbf{d} indicates the Michaelis-Menten half-saturation constant, $c_{\rm g}^*$.  We find that in the single colony case, the carbon dioxide concentration rapidly approaches and exceeds saturation at the location of the colony. The glucose profiles show strong gradient with only a small fraction of cells at the colony periphery experiencing concentrations above $c_{\rm g}^*$ (solid gray line in panel \textbf{b}). The simulation of two neighboring colonies illustrate the higher CO$_2$, and lower glucose concentration in the region between the two colonies. This observation is consistent with experimental measurement in Movie S6, and the deviation of bubbles towards each other (Fig. 5). 
        }
        \label{fig:RDSim}
    \end{figure}

Fig.~S\ref{fig:RDSim}\textbf{a} and \textbf{b} respectively show the carbon dioxide and glucose profiles at different time points. We find that glucose is rapidly depleted due to uptake by the colony, diffusing only a short distance into the periphery of the colony. Consequently, only cells at the edge of the colony access nutrients and actively ferment, producing CO$_2$. Nonetheless, the CO$_2$ concentrations quickly reach saturation near the center of the colony.

We then investigate the interaction between the CO$_2$ (Fig.~S\ref{fig:RDSim}\textbf{c}) and the glucose (Fig.~S\ref{fig:RDSim}\textbf{d}) fields produced by two neighboring colonies. We extend the model to a cylindrical symmetrical model  and simulate two yeast colonies separated by 10 mm, initially in a uniform glucose environment. As in the single-colony case, we observe rapid saturation of CO$_2$ levels near the two colonies. Notably, the CO$_2$  concentration peaks in the region between the two colonies. This spatial accumulation of CO$_2$ reduces the local pH and thus softens the matrix in the intercolony region, biasing bubble trajectories toward the neighboring colony---consistent with the experimentally observed deviation angles (Fig. 5 of main text).

\begin{table}
\begin{tabular}{p{3.2cm}p{4.5cm}p{4cm}p{2cm}}
 \hline 
 \hline
Physical parameters & Definition & Range of values & Reference \\
 \hline
 $D_{\rm g}$ & Glucose diffusion coefficient & 2.8 mm$^2$ h$^{-1}$ & Ref.~\citep{martinez2022morphological} \\
 $\kappa$ & Maximum glucose consumption rate & $2 \times 10^{-2}$ $\mu$mol (g h)$^{-1}$ & Refs.~\citep{fiechter1992regulation, fiechter1981regulation} \\
 $c_{\rm g}^*$ & Michaelis Menten half saturation coefficient & $5.55 \times 10^{-10}$ mol mm$^{-3}$ & Ref.~\citep{fink2023microbial} \\
$c_{\rm g, 0}$ & Far field glucose concentration & $1.1 \times 10^{-7}$ mol mm$^{-3}$ & Experimental \\
  $D_{\rm CO_2}$ & CO$_2$ diffusivity coefficient & 6.84 mm$^2$ h$^{-1}$ & Ref.~\citep{polat2024diffusivity} \\
 $\alpha$ & Maximal CO$_2$ production rate & $0.05$  mol (g h)$^{-1}$ & Ref.~\cite{paciello2014fermentative} \\
 $\rho$ & Close packed cell density & $1 \times 10^{-3}$ g mm$^{-3}$ & Estimated \\

 \hline

 \end{tabular}
 \caption{Estimates of the physical parameters used in the reaction diffusion simulation. \label{tab:PenLength}}
 \end{table}

\subsection{Simulations of Darwin's drift modified for yield stress fluids}

\subsubsection{Numerical Simulations in Basilisk}
    We perform direct numerical simulations (DNS) of a rising bubble in a viscoplastic (Bingham) fluid using Basilisk C.
    Basilisk C is an open-source language developed by S. Popinet and collaborators \cite{Popinet2003, Popinet2009, Popinet2015} for solving differential equations on adaptive Cartesian meshes. 
    In this work, we use the built-in Navier–Stokes and Volume-of-Fluid (VOF) solvers~\cite{popinet2018numerical}, which have been extensively validated in previous studies [see \cite{Popinet2009} for detailed comparisons], together with the implementation of an inelastic viscoplastic constitutive model (Bingham), as previously used in a series of Plastocapillarity problems~\cite{jalaal2021spreading,francca2024elasto} including bubble dynamics~\cite{jalaal2016long,Sanjay_Lohse_Jalaal_2021,zare2024bubble, esposito2025rising}.
    In addition, to characterize the deformation of the matrix we construct a dynamic grid of Lagrangian tracer particles that are advected using the Runge-Kutta scheme implemented in \cite{Antoonvh_tracer_particles_2026}.
    The simulations are carried out assuming axisymmetry to reduce computational cost.

\subsubsection{Initial configuration}
    \begin{figure}[h]
        \centering
        \includegraphics[width=0.5\textwidth]{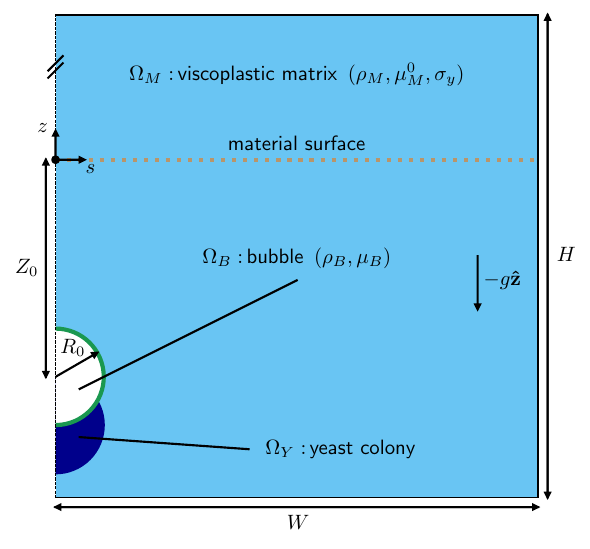}
        \caption
        {
            The initial configuration of the DNS for a single bubble rising through a viscoplastic material. 
            Shown are the two material regions $\Omega_M$ and $\Omega_B$ corresponding to the matrix and the bubble respectively.
            The matrix is a viscoplastic (Bingham) fluid with density $\rho_M$, plastic viscosity $\mu_M^0$, and yield stress $\sigma_y$.
            The bubble contains air with density $\rho_B$ and viscosity $\mu_B$.
            Shown also is an arrow representing the constant gravitational acceleration $g$.
            The domain is an axisymmetric cylinder with height $H$ and width $W$.
            The bubble is initially spherical and has radius $R_0$.
        }
        \label{fig:DNS_single_bubble}
    \end{figure}
    The initialization of the DNS for single bubble is shown in Fig.~\ref{fig:DNS_single_bubble}.
    The domain $\Omega$ is assumed to be axisymmetric and is parametrized using the cylindrical coordinates $(s, z)$.
    The fluid domain $\Omega$ is divided into two disjoint material regions, $\Omega_M$ and $\Omega_B$ representing the matrix and bubble respectively.
    The fluid in $\Omega_M$ is viscoplastic with mass density $\rho_M$, plastic viscosity $\mu_M^0$ and yield stress $\sigma_y$.
    The fluid in $\Omega_B$ is Newtonian with mass density $\rho_B$ and viscosity $\mu_B$.
    The system is initialized with $\Omega_B$ a sphere of radius $R_0$ centered on the point $(0, -Z_0)$.
    An additional region $\Omega_Y$ represents the yeast colony entrained by the bubble, and is initialized as the intersection of the region $\Omega_M$ with a sphere radius $R_0$ centered on the point $(0, -Z_0 - R_0)$.
    The boundaries of $\Omega$ are the surfaces $z = \pm H/2$ and $s = W/2$ where $H$ is the height of the domain and $W$ is the width.
    The velocity $\bm{u} = (u_s, u_z)$ is initialized with $\bm{u} = \bm{0}$ everywhere and the pressure $p$ is initialized with $p = 0$ everywhere.
    The time-evolution of the regions $\Omega_M$, $\Omega_B$ and $\Omega_Y$ as well as the flow fields $p$ and $\bm{u}$ is discussed in the following section.

\subsubsection{Equations of motion}
    The equations of motion are the continuity equation and the generalized Navier--Stokes momentum equation:
    \begin{align}
        \bm{\nabla} \cdot \bm{u} &= 0 
        \label{eqn:continuity_appendix} 
        \\
        \rho \left( \pdv{\bm{u}}{t} + \bm{u}\cdot\bm{\nabla u} \right)
        &=
        -\bm{\nabla} p 
        + 
        \bm{\nabla} \cdot \bm{\sigma}
        +
        \bm{f}_\gamma
        +
        \bm{f}_B.
        \label{eqn:momentum_appendix} 
    \end{align}
    where $\rho$ is the mass density, $\bm{u}$ is the velocity field, $p$ is the pressure field, and $\bm{\sigma}$ is the deviatoric stress tensor.
    We adopt the generalized--Newtonian formalism, in which $\bm{\sigma} = 2\mu\bm{\mathcal{D}}$ where $\bm{\mathcal{D}} = (1/2)(\bm{\nabla} \bm{u} + [\bm{\nabla} \bm{u}]^T)$ is the rate-of-strain tensor and the viscosity $\mu$ depends on $\bm{\mathcal{D}}$ in order to accommodate the non-Newtonian fluid behavior.
    The surface tension force $\bm{f}_\gamma$ is written as
    $\bm{f}_\gamma = \gamma \kappa \bm{\hat{n}} \delta_S$,
    where $\gamma$ is the surface tension coefficient, $\kappa$ is the mean curvature of the interface, $\bm{\hat{n}}$ its unit normal, and $\delta_S$ a Dirac delta distribution restricting the force to the interface $S = \partial\Omega_B$.
    The external body force $\bm{f}_B$ is written as $\bm{f}_B = -\rho g \bm{\hat{z}}$ and gives rise to the buoyancy of the bubble.
    The disparate material properties of the $\Omega_B$ and $\Omega_M$ are encoded using a scalar color function $c(\bm{x},t)$, which takes the value 0 inside $\Omega_B$ and 1 inside $\Omega_M$. 
    Thus $\rho$ and $\mu$ may be written as:
    \begin{align}
        \rho  &= \rho_M c(\bm{x},t) + \rho_B[1 - c(\bm{x},t)] \\
        \mu   &= \mu_M c(\bm{x},t) + \mu_B[1 - c(\bm{x},t)].
    \end{align}
    where $\mu_M$ is the effective viscosity of the viscoplastic material given by:
    \begin{align}
        \mu_M
        &=
        \mu_M^0
        +
        \frac{\sigma_y}{\norm{\bm{\mathcal{D}}} + \epsilon}
        \left(
            1
            -
            e^{-M \norm{\bm{\mathcal{D}}} }
        \right) 
        \label{eqn:app_visc_appendix} 
    \end{align}
    where $\norm{\bm{\mathcal{D}}} = \sqrt{(1/2)\,\bm{\mathcal{D}} :\bm{\mathcal{D}}}$ is the second invariant of $\bm{\mathcal{D}}$.
    The Papanastasiou regularisation parameter, $M$, is included such that the effective viscosity of the matrix is smooth and well-defined in the limit 
    $\norm{\bm{\mathcal{D}}} \to 0$; a condition necessary for the stability of the simulation \cite{Papanastasiou}.
    In the limit $M \to \infty$, we recover the Bingham model.
    We choose $M$ to be large enough such that our numerical solutions are sufficiently close to the Bingham model.
    In addition, we include a small parameter $\epsilon$ in order to avoid zero-division errors. 
    Note that the inclusion of $\epsilon$ does not affect the limiting behavior of the stress tensor in the limit $\norm{\bm{\mathcal{D}}} \to 0$.
    Equations~\ref{eqn:continuity_appendix},~\ref{eqn:momentum_appendix} and~\ref{eqn:app_visc_appendix} are non-dimensionalized by choosing the following characteristic length, pressure and velocity scales:
    \begin{align}
        L_0 &= R_0,\qquad &
        P_0 &= \Delta\rho\,g\,R_0,\qquad &
        U_0 &= \frac{\Delta\rho\,g\,R_0^{2}}{\mu_M^{0}} .
    \end{align}
    where $\Delta\rho = \rho_M - \rho_B$ is the density difference between the matrix and the bubble.
    This choice of scales gives rise to the following dimensionless parameters:
    \begin{align}
        \text{Ar}_0
        &=
        \frac{\rho_M \Delta\rho \, g R_0^{3}}{(\mu_M^0)^2}
        &
        \text{Bo}_0
        &=
        \frac{\Delta \rho \, gR_0^2}{\gamma}
        &
        \text{Bi}_0
        &=
        \frac{\sigma_y}{2\Delta \rho\, gR_0}
    \end{align}
    known as the Archimedes number, the Bond number and the Bingham number respectively.
    Note that an arbitrary factor of $2$ in the definition $\text{Bi}_0$ is included for consistency with the experiments.
    The subscript $_0$ is included to distinguish the initialization values from their time-dependent counterparts.
    In addition we define the density and viscosity ratios 
    $\alpha = \rho_B/\rho_M$ and $\lambda = \mu_B/\mu_M^0$ respectively.
    In dimensionless form, the equations of motion then read:
    \begin{align}
        \bar{\bm{\nabla}} \cdot \bar{\bm{u}} &= 0
        \label{eqn:continuity_nondim_appendix}
        \\
        \text{Ar}_0 \bar{\rho}
        \left(\pdv{\bar{\bm{u}}}{\bar{t}} + \bar{\bm{u}} \cdot \bar{\bm{\nabla}} \bar{\bm{u}}\right)
        &=
        -\bar{\bm{\nabla}} \bar{p}
        +
        \bar{\bm{\nabla}} \cdot (2\bar{\mu}\bar{\bm{E}})
        +
        \frac{1}{\text{Bo}_0}
        \bar{\kappa} \bm{\hat{n}} \bar{\delta}_S
        -
        \frac{1}{1-\alpha}
        \bar{\rho}
        \bm{\hat{z}}
        \label{eqn:momentum_nondim_appendix}
    \end{align}
    where 
    \begin{align}
        \bar{\rho} &= c(\bm{x},t) + \alpha [1 - c(\bm{x},t)] \\
        \bar{\mu} 
        &= 
        \bar{\mu}_M
        c(\bm{x},t) 
        + 
        \lambda
        [1 - c(\bm{x},t)].
    \end{align}
    and
    \begin{align}
        \bar{\mu}_M
        &=
        1
        +
        \frac{2\text{Bi}_0}{\norm{\bm{\bar{\mathcal{D}}}} + \bar{\epsilon}}
        \left(
            1
            -
            e^{-\bar{M} \norm{\bar{\bm{\mathcal{D}}}} }
        \right) 
        \label{eqn:app_visc_nondim_appendix} 
    \end{align}
    Note that the $\,\bar{}\,$ symbol is used to indicate that a quantity is dimensionless.
    For each simulation, $\bar{\epsilon}= 10^{-20}$ and $\bar{M} = 10^6$.
    The yeast colony is implemented as its own scalar color function $d(\bm{x}, t)$, which takes the value 0 inside $\Omega_Y$ and 1 outside $\Omega_Y$.
    The time evolution of $\Omega_M$, $\Omega_B$ and $\Omega_Y$ is solved by advecting $c(\bm{x}, t)$ and $d(\bm{x}, t)$ using the VOF method.
    In each simulation, given that the inertia and viscosity of the bubble have negligible effect on the dynamics, we fix $\alpha = \lambda = 0.01 \ll 1$.
    In general, inertial effects are negligible, thus, unless otherwise stated, we choose $\text{Ar}_0 = 0.1 \ll 1$.
    Moreover, $H/R_0$ and $W/R_0$ are chosen to be large enough that the boundary conditions do not affect the dynamics.
    For simulations involving a single bubble, $H/R_0 = W/R_0 = 40$ with $Z_0/R_0 = 15$.
    For simulations involving multiple bubbles, $H/R_0 = W/R_0 = 100$ with identical bubbles located at $(0,-Z_0), (0,-Z_0 - D_0), (0,-Z_0 - 2D_0)$ and $(0,-Z_0 - 3D_0)$, where the bubble spacing $D_0 = 10$.
    The maximum level of mesh refinement, \texttt{LEVEL}, is defined such that the domain size $L = \max\{W, H\}$ is $2^{\texttt{LEVEL}}$ times the size of the smallest cell.
    Consequently, the maximum number of cells across the initial bubble diameter is
    $2^{\texttt{LEVEL}} \cdot 2R_0/L$.
    The value of \texttt{LEVEL} is chosen to guarantee a minimum of 40 cells across the diameter of the bubble.
    In particular, for single-bubble simulations $\texttt{LEVEL} = 10$ and for multiple-bubble simulations $\texttt{LEVEL} = 11$.
    
\subsection{The critical Bingham number}
\begin{figure}[h]
        \includegraphics[width=0.5\textwidth]{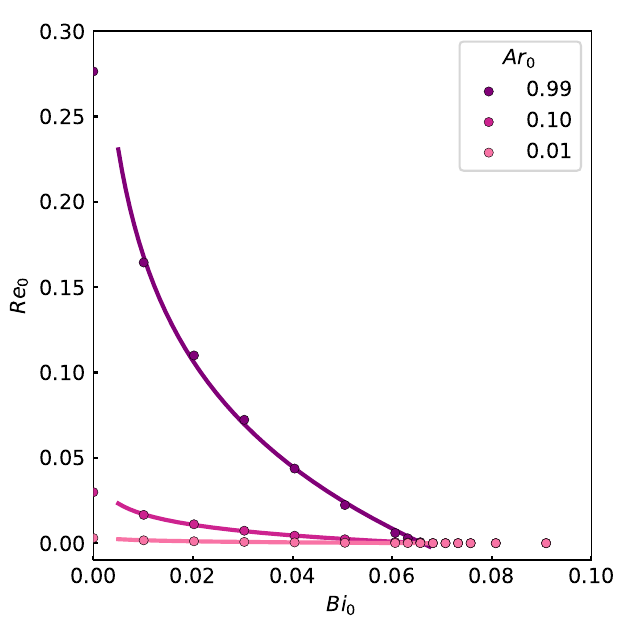}
        \includegraphics[width=0.5\textwidth]{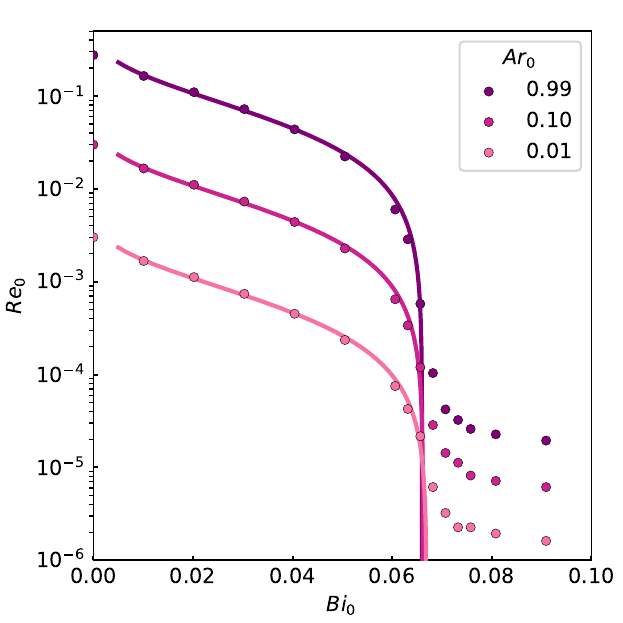}
        \caption{
            The terminal Reynolds number, $\text{Re}_0$, for single rising bubbles at different $\text{Ar}_0$, and $\text{Bi}_0$. 
            Here, $\text{Bo}_0 = 1$ for all simulations. 
            The dots indicate simulations results, while the full line indicates the logarithmic fit. 
            Using the logarithmic fit, we estimate the critical Bingham number to be $\text{Bi}_{c} \approx 0.066$, independently of $\text{Ar}_0$ in the range $0.01 < \text{Ar}_0 < 1.0$.
        }
        \label{fig:critical_bingham_number}
    \end{figure}

    Here, we discuss numerical estimates of the critical value, $\text{Bi}_c$, of $\text{Bi}_0$, above which the buoyancy is insufficient to yield the material and the bubble remains trapped.
    The terminal rise speed of the bubble, $U_b$, is used to construct a Reynolds number, $\text{Re}_0$, defined as:
    \begin{align}
        \text{Re}_0
        &=
        \frac{\Delta \rho \, U_b R_0}{\mu_M^0}
    \end{align}
    where the subscript $_0$ indicates that $\text{Re}_0$ is constructed using the initial radius $R_0$, rather than the time-dependent $R(t)$, and the matrix plastic viscosity $\mu_M^0$ rather than the effective viscosity $\mu_M$.
    Fig.~\ref{fig:critical_bingham_number} shows $\text{Re}_0$ as a function of $\text{Bi}_0$ in the range $0 < \text{Bi}_0 < 0.1$, for different values of $\text{Ar}_0$ in the range $0.01 < \text{Ar}_0 < 1.0$.
    The value of $\text{Bi}_c$ is then defined as the least value of $\text{Bi}_0$ for which $\text{Re}_0 = 0$.
    Since the yielding transition is regularized, $\text{Re}_0 > 0$ for all $\text{Bi}_0$, thus $\text{Bi}_c$ is not well-defined.
    However, $\text{Bi}_c$ can be estimated by logarithmically fitting the data in the range $0 < \text{Bi}_0 < 0.06$ where the bubble clearly rises, and then extrapolating.
    This gives rise to an estimate $\text{Bi}_{\text{c}} \approx 0.066$ independently of $\text{Ar}_0$ in the range $0.01 < \text{Ar}_0 < 1.0$.
    This value is in approximate agreement with \cite{tsamopoulos2008steady} in which it is found that $\text{Bi}_{\text{c}} \approx 0.0725$.
    
\subsection{Effect of Bingham number on entrainment}
    \begin{figure}[h]
        \includegraphics[width=1\textwidth]{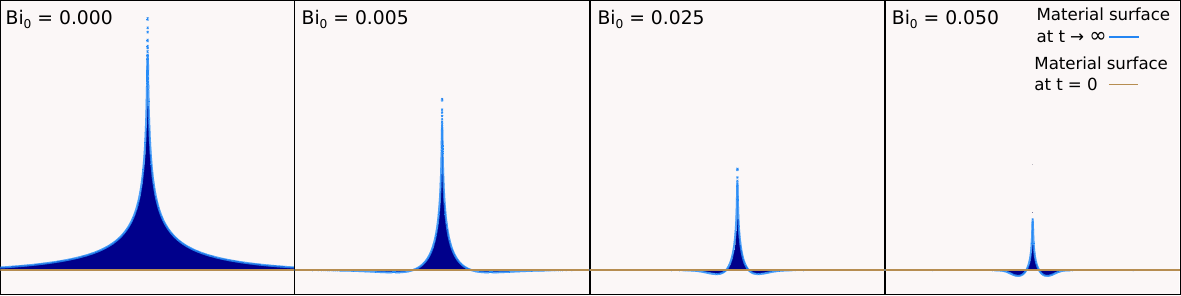}
        \caption{
            The material entrainment surface at large times $\bar{t}\to\infty$, for single rising bubbles with different values of the Bingham number $\text{Bi}_0$, in the range $0.0 < \text{Bi}_0 < 0.05$ with $\text{Ar}_0 = 0.1$ and $\text{Bo}_0 = 1.0$ fixed.
            The forward entrainment decreases with increasing $\text{Bi}_0$.
        }
        \label{fig:entrainment_vs_Bi}
    \end{figure}

    Fig.~\ref{fig:entrainment_vs_Bi} shows the effect of $\text{Bi}_0$ on the entrainment by the rising bubble.
    In each simulation, $\text{Ar}_0 = 0.1$ and $\text{Bo}_0 = 1.0$ are fixed and $\text{Bi}_0$ is varied across the range $0.0 < \text{Bi}_0 < 0.1$.
    Mass conservation forces the net entrainment, that is the sum of the forward and backward entrainment, to be zero.
    The forward entrainment decreases with increasing $\text{Bi}_0$.
    For $\text{Bi}_0 > \text{Bi}_c$, the bubble remains fixed in place and the forward entrainment is zero.

\clearpage
\section{Extended data}\label{sec3}

    \begin{figure}[h]
        \includegraphics[width=1\textwidth]{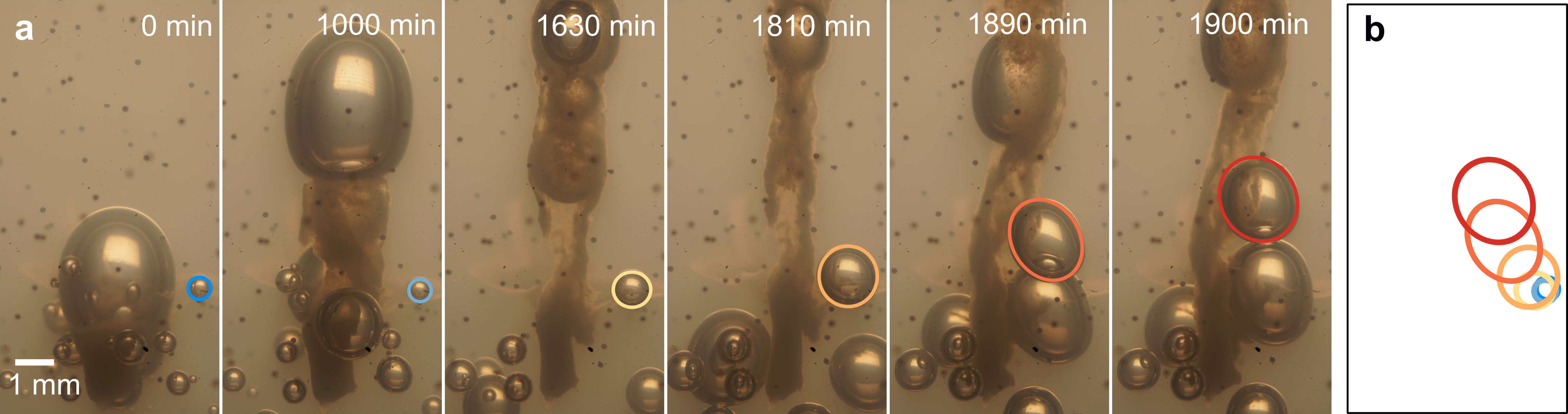}
        \caption{ \textbf{a,} The time series of the growth of a bubble, shown by a solid line, near the columnar colony. The bubble grow towards the colony suggesting that there is a gradient in the yield stress values. The bubble growth is deviated toward the softer region (closer to the colony). \textbf{b,} The superposition of the bubble perimeter obtained from the experiment in panel \textbf{a}. The passage of multiple other bubbles during the the bubble growth shows that the biased growth toward the colony is due to the gradient in the yield stress and not merely the bubble-bubble hydrodynamic interactions.
        }
        \label{fig:Biased_growth}
    \end{figure}

 \begin{figure}[h]
          \centering
        \includegraphics[width=0.5\textwidth]{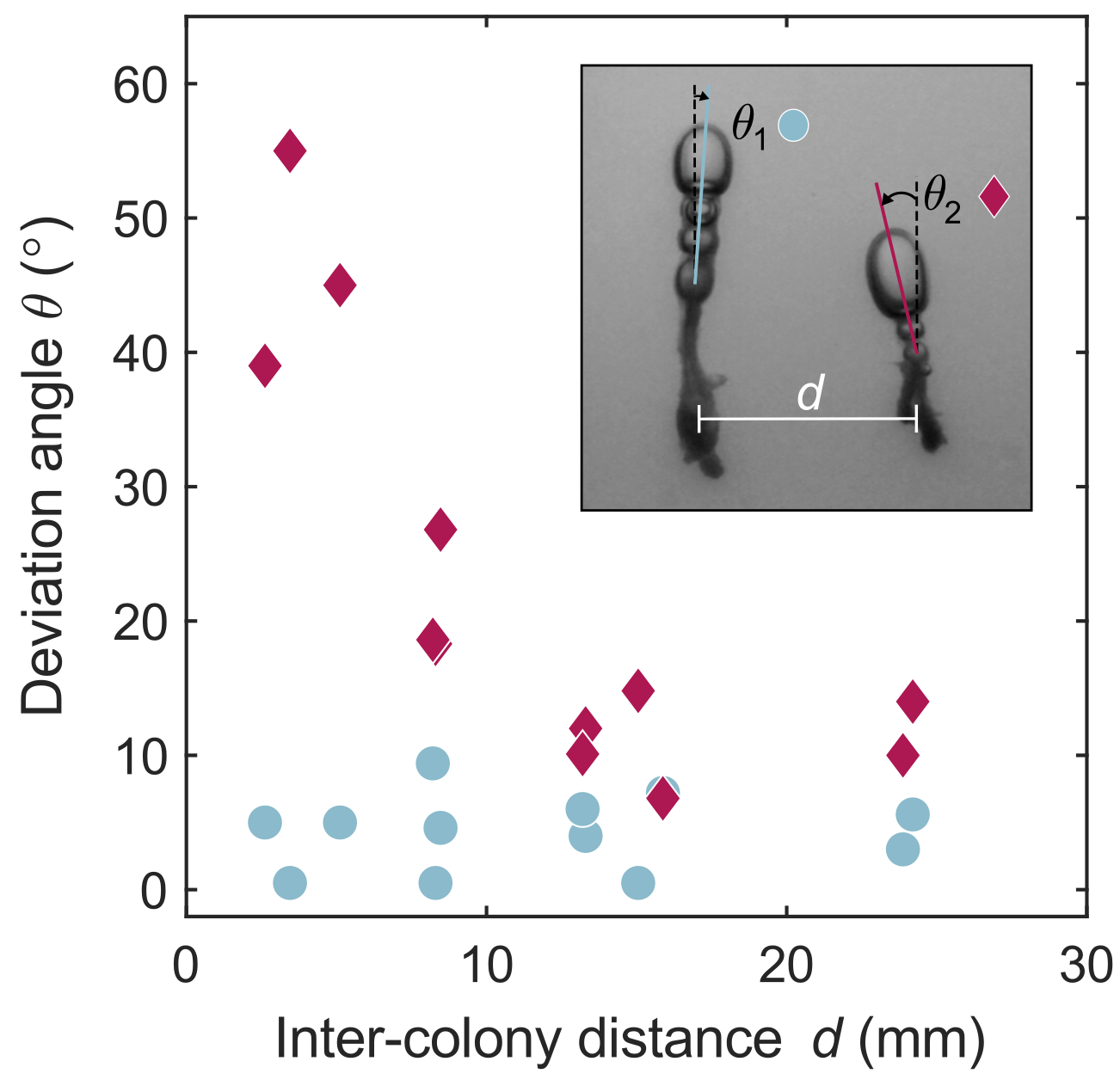}
        \caption{ Variation of the bubble deviation angles with increasing inter-colony distance. The bubble at the lower height experiences a larger deviation angle. 
        }
        \label{fig:Theta}
    \end{figure}

\clearpage

\end{document}